\documentclass[longauth]{aa}  

\usepackage{graphicx}
\usepackage{natbib}
\usepackage{txfonts}
\newcommand{\kms}{\ensuremath{{\rm km\, s^{-1}}}}
\begin{document}

   \title{GRAVITY chromatic imaging of $\eta$ Car's core}

\subtitle{milliarcsecond resolution imaging of the wind-wind collision zone (Br$\gamma$, He {\sc i})} 

   \author{GRAVITY Collaboration\thanks{GRAVITY is developed in a
       collaboration by the Max Planck Institute for Extraterrestrial Physics,
LESIA of Paris Observatory and IPAG of Université Grenoble Alpes / CNRS,
the Max Planck Institute for Astronomy, the University of Cologne, the
Centro Multidisciplinar de Astrofisica Lisbon and Porto, and the European
Southern Observatory.}: J. Sanchez-Bermudez\inst{1,8} \and G.~Weigelt \inst{11} \and
J. M.~Bestenlehner\inst{1,15} \and P.~Kervella
\inst{2,13} \and
W.~Brandner\inst{1} \and Th.~Henning\inst{1}  \and A.~Müller \inst{1} \and G.~Perrin \inst{2} \and
J.-U.~Pott  \inst{1} \and M.~Schöller \inst{7} \and R.~van Boekel
% ...
}

   \institute{Max Planck Institute for Astronomy, K\"{o}nigstuhl 17,
     Heidelberg, Germany, D-69117 \\
  \email{jsanchez@mpia.de}
\and
LESIA, Observatoire de Paris, PSL Research University, CNRS, Sorbonne Universit\'es, UPMC Univ. Paris 06, Univ. Paris Diderot, Sorbonne Paris Cit\'e, France
             \and 
             Univ. Grenoble Alpes, CNRS, IPAG, F-38000 Grenoble, France
             \and
             I. Physikalisches Institut, Universität zu Köln, Zülpicher Str. 77, 50937, Köln, Germany
             \and
Universidade do Porto - Faculdade de Engenharia, Rua Dr. Roberto Frias, 4200-465 Porto, Portugal
\and
CENTRA, Instituto Superior Tecnico, Av. Rovisco Pais, 1049-001 Lisboa, Portugal
\and 
European Southern Observatory, Karl-Schwarzschild-Str. 2, 85748
Garching, Germany
\and
European Southern Observatory, Casilla 19001, Santiago 19, Chile
\and
Max Planck Institute for Extraterrestrial Physics, Giessenbachstrasse, 85741 Garching bei M\"{u}nchen, Germany
\and
             Dublin Institute for Advanced Studies, 31 Fitzwilliam Place, D02\,XF86 Dublin, Ireland
\and
Max-Planck-Institute for Radio Astronomy, Auf dem H\"ugel 69,
53121 Bonn, Germany
\and 
Department of Physics, Le Conte Hall, University of California, Berkeley, CA 94720, USA
\and 
Unidad Mixta Internacional Franco-Chilena de Astronomía (CNRS UMI 3386), Departamento de Astronomía, Universidad de Chile, Camino El Observatorio 1515, Las Condes, Santiago, Chile
\and 
Sterrewacht Leiden, Leiden University, Postbus 9513, 2300 RA Leiden,
The Netherlands
\and
Department of Physics and Astronomy, University of Sheffield, Hicks
Building, Hounsfield Rd, Sheffield, S3 7RH, UK
% ...
}

   \date{Received ; accepted }

\titlerunning{$\eta$ Car Chromatic Imaging}
 
  \abstract
{$\eta$ Car is one of the most intriguing luminous blue variables in
  the Galaxy. Observations and models of the X-ray, ultraviolet, optical, and infrared
  emission suggest a central binary in a highly eccentric orbit with a
  5.54 yr period residing in its core. 2D and 3D radiative transfer and hydrodynamic
  simulations predict a primary with
  a dense and slow stellar wind that interacts with
  the faster and lower density wind of the secondary. The
  wind-wind collision scenario suggests that the secondary's wind
  penetrates the primary's wind creating a low-density cavity in it,
  with dense walls where the two winds interact. However, the
  morphology of the cavity and its physical properties are not yet
  fully constrained.}
   {We aim to trace the inner $\sim$5--50 au structure of $\eta$ Car's
     wind-wind interaction, as seen through Br$\gamma$ and, for
     the first time, through the He {\sc
       i} 2s-2p line.
     }
   {We have used spectro-interferometric observations with
     the $K$-band beam-combiner GRAVITY at the VLTI. The analyses of
     the data include (i) parametrical model-fitting to the
     interferometric observables, (ii) a \texttt{CMFGEN} model of the source's
     spectrum, and (iii) interferometric image reconstruction. }
   {Our geometrical modeling of the continuum data allows us to estimate its
     FWHM angular size close to 2 mas and an elongation ratio
     $\epsilon$ = 1.06 $\pm$ 0.05 over a PA =
     130$^{\circ}$ $\pm$ 20$^{\circ}$. Our \texttt{CMFGEN} modeling of the spectrum helped us
     to confirm that the role of the secondary
     should be taken into account to properly reproduce the observed
     Br$\gamma$ and He {\sc i} lines. Chromatic images across the
     Br$\gamma$ line reveal a southeast arc-like feature, possibly associated
     to the hot post-shocked winds flowing along the cavity wall. The images of the He {\sc i} 2s-2p line served to constrain
   the 20 mas ($\sim$ 50 au) structure of the line-emitting
   region. The observed morphology of He {\sc i} suggests that the secondary is responsible for the ionized
   material that produces the line profile. Both the Br$\gamma$ and the He {\sc i} 2s-2p
   maps are consistent with previous hydrodynamical models of the
   colliding wind scenario. Future dedicated simulations
 together with an extensive interferometric campaign are necessary to refine our
 constraints on the wind and stellar parameters of the binary, which finally will
 help us predict the evolutionary path of $\eta$ Car.}
   {}


   \maketitle
%

\begin{figure*}[htp]
\centering
\includegraphics[width=16 cm]{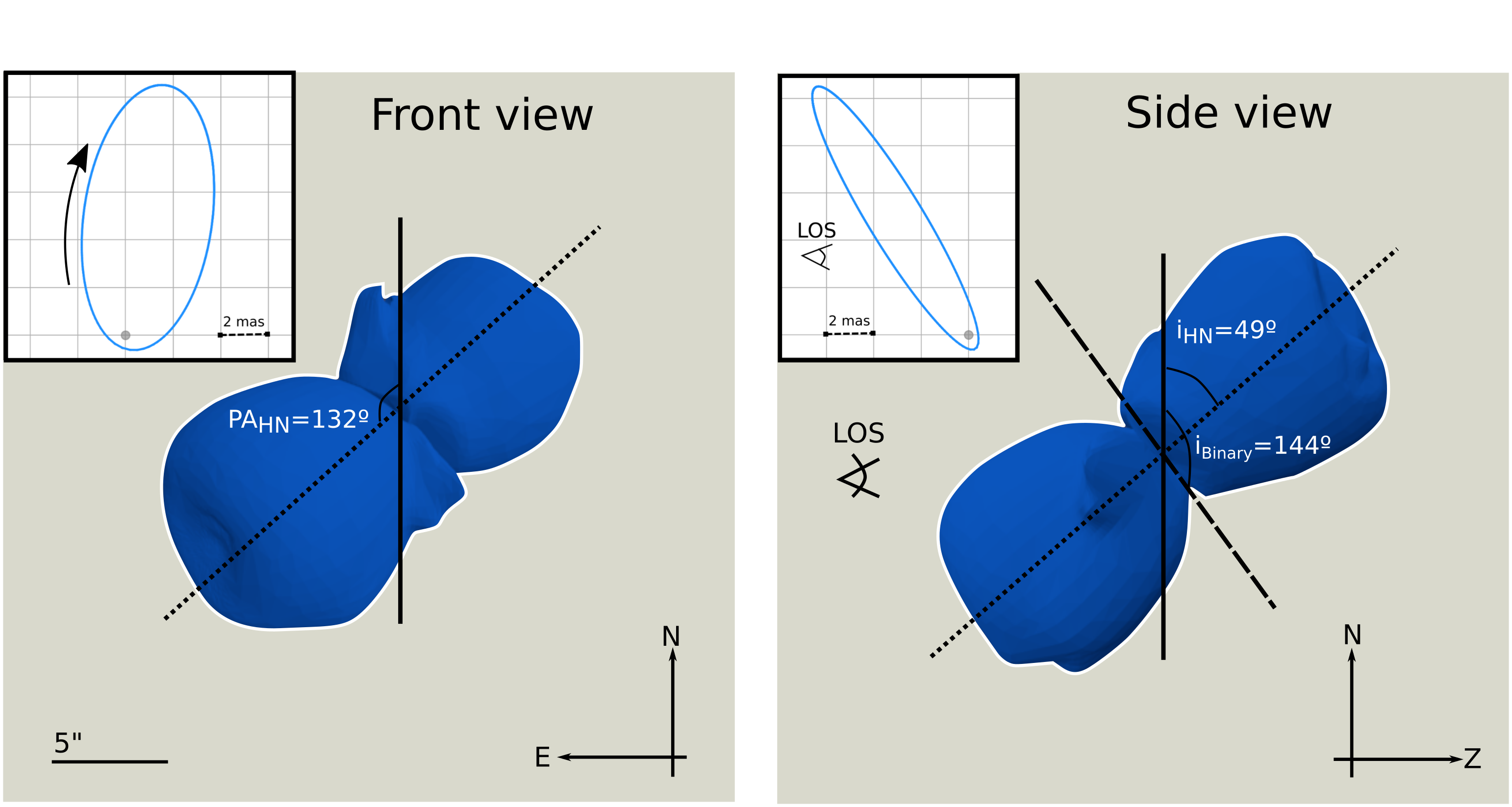}
\caption{Homunculus Nebula and the binary orbit
  projected in the plane of the sky (\textit{left panel}; Front view) and in a
  plane formed by the observer's LOS and the sky plane
  (\textit{right panel}; Side view). The
3D Homunculus model was obtained from the online resources of \citet{Steffen_2014}.}
\label{fig:EtaCar_orientation} 
\end{figure*}

\section{Introduction}

Massive stars are among the most important chemical factories of the
interstellar medium (ISM). This is mainly because their evolution and
fate are highly affected by strong stellar winds ($v_{\rm{w}} \sim$10$^3~\kms$), high mass-loss rates ($\dot{M} \sim$ 10$^{-5}$--10$^{-3}$
$M_{\odot}$ yr$^{-1}$) and deaths as supernovae (SNe) \citep[see, e.g., ][]{Conti_1976, Langer_1994,
  Meynet_2003, Meynet_2011}. One of the evolutionary stages of high-mass stars
that exhibit sporadic but violent mass-loss episodes is the luminous blue
variable phase \citep[LBV; ][]{Humphreys_1994}. The importance of LBVs
to stellar evolution models relies on the possibility of them to
directly explode as SN without being a Wolf-Rayet (WR) star \citep{Smith_2007, Smith_2011, Trundle_2008}. Therefore, detailed studies of LBVs are crucial to understand the mass-loss
processes in high-mass stars \citep[see e.g.,
][]{Pastorello_2010, Smith_2011}.

The source, $\eta$ Car, is one of the most intriguing LBVs in the Galaxy. Located
at the core of the Homunculus Nebula in the Trumpler 16 cluster
at a distance of 2.3 $\pm$ 0.1 kpc \citep{Walborn_1973, Allen_1993,
  Smith_2006}, it has been identified as a luminous \citep[$L_{\mathrm{tot}}
\ge$ 5x10$^6$ $L_{\odot}$][]{Davidson_1997, Smith_2003} colliding-wind
binary \citep{Damineli_1997, Hillier_2001, Damineli_2008a,
  Damineli_2008b, Corcoran_2010} in a highly
eccentric orbit \citep[e $\sim$ 0.9; ][]{Corcoran_2005} with a period of 2022.7 $\pm$
1.3 d \citep{Damineli_2008b}. 

The primary, $\eta_A$, is a very
massive star with an estimated  $M \sim$ 100 $M_{\odot}$, a mass-loss rate $\dot{M}
\sim$ 8.5 x 10$^{-4}$ $M_{\odot}$ yr$^{-1}$ and a wind terminal speed $v_{\infty}
\sim$ 420$~\kms$  \citep{Hillier_2001, Hillier_2006, Groh_2012a,
  Groh_2012b}. Evidence
suggests that $\eta_A$ is near the Eddington limit \citep{Conti_1984, Humphreys_1994}. Therefore, it loses
substantial mass in violent episodes such as the ``Great Eruption''
where $\sim$15 $M_{\odot}$, possibly more than 40 $M_{\odot}$
\citep{Gomez_2010, Morris_2017}, were ejected over a
period of $\sim$10 yr. Mass-loss events dominate the
evolutionary tracks of the most massive stars \citep{Langer_1998,
  Smith_2006}. In binary star systems, like
$\eta$ Car, the presence of a companion affects wind-driven mass
loss, providing alternative evolutionary pathways compared to single
stars. Therefore, understanding in detail their mass-loss processes is
particularly important. 
        
The nature of the secondary, $\eta_B$, is even less constrained since it has
not been directly observed \citep[it is, at least, of the order of 100 times fainter than
$\eta_A$;][]{Weigelt_2007} because it is embedded in the dense wind
of the primary.  Models of the X-ray emission (kT $\sim$ 4--5 keV) predict a
wind terminal velocity $v_{\infty}
\sim$ 3000$~\kms$ , a mass-loss rate $\dot{M} \sim$
10$^{-5}~M_{\odot}$ yr$^{-1}$ \citep{Pittard_2002, Okazaki_2008, Parkin_2011b}, and a T$_{\mathrm{eff}} \sim$ 36000--41000 K
\citep{Teodoro_2008}. 

Models of the mutual motion of the stars
suggest an inclination $i \sim$ 130--145$^{\circ}$, an
argument of periastron $\omega \sim$ 240--285$^{\circ}$,
and a sky projected PA $\sim$ 302--327$^{\circ}$ for the best-fit orbital solution \citep{Damineli_1997, Okazaki_2008, Parkin_2009,
  Parkin_2011b, Groh_2010b, Gull_2011, Madura_2012a, Madura_2012b,
  Teodoro_2016}. This suggests that the orbital plane of the
binary is almost perpendicular to the Homunculus axis \citep[ $i \sim$
49$^{\circ}$ with respect to the sky plane; PA$\sim$132$^{\circ}$;][]{Davidson_2001,
  Smith_2006}. These orbital parameters also imply that the secondary
remains in front of the primary (in the line-of-sight -LOS- of the observer)
most of the time during the orbital motion, except close to the
periastron, where $\eta_B$ goes behind $\eta_A$ and it is obscured by
the dense primary wind.

Figure\,\ref{fig:EtaCar_orientation} displays the inclination and
PA of the Homunculus Nebula and of the orbit of $\eta_B$
around $\eta_A$,  according to the orbital solution reported by
\citet{Teodoro_2016}. The left panel displays the nebula
projected in the plane of the sky. The PA (east to the
north) of the semi-major axis is labeled in the image. The right panel displays
the position of the nebula and of the orbit of the binary in a plane parallel to the
observer's line-of-sight and the sky plane. The inclination angles
(relative to the sky plane) of the nebula ($i_{\mathrm{HN}}$) and of the binary's orbital
plane ($i_{\mathrm{Binary}}$) are
labeled.

The secondary, $\eta_B$, photoionizes part of the primary wind,
changing the strength of lines such as H$\alpha$, He {\sc i}, [Fe
{\sc ii}], and [Ne {\sc ii}]
\citep{Hillier_2001, Nielsen_2007, Mehner_2010, Mehner_2012,
  Madura_2012b, Davidson_2015}. Additionally, it ionizes the inner 1'' circumstellar
ejecta \citep[][]{Weigelt_1986, Hofmann_1988, Weigelt_1995}. 2D radiative
transfer models and 3D hydrodynamical simulations of the wind-wind
collision scenario suggest that the high-velocity secondary wind penetrates the slower and denser primary wind creating a
low-density cavity in it, with thin and dense walls where
the two winds interact \citep{Okazaki_2008, Groh_2012a, Madura_2012b,
Madura_2013, Clementel_2015a, Clementel_2015b}. This wind-wind
collision scenario produces the shock-heated gas responsible for the X-ray
variability, and the ionization effects observed in the 
optical, infrared, and ultraviolet spectra.

\begin{figure}[thp]
\centering
\begin{minipage}[c]{0.4\textwidth}
\centering
    \includegraphics[width = 8 cm]{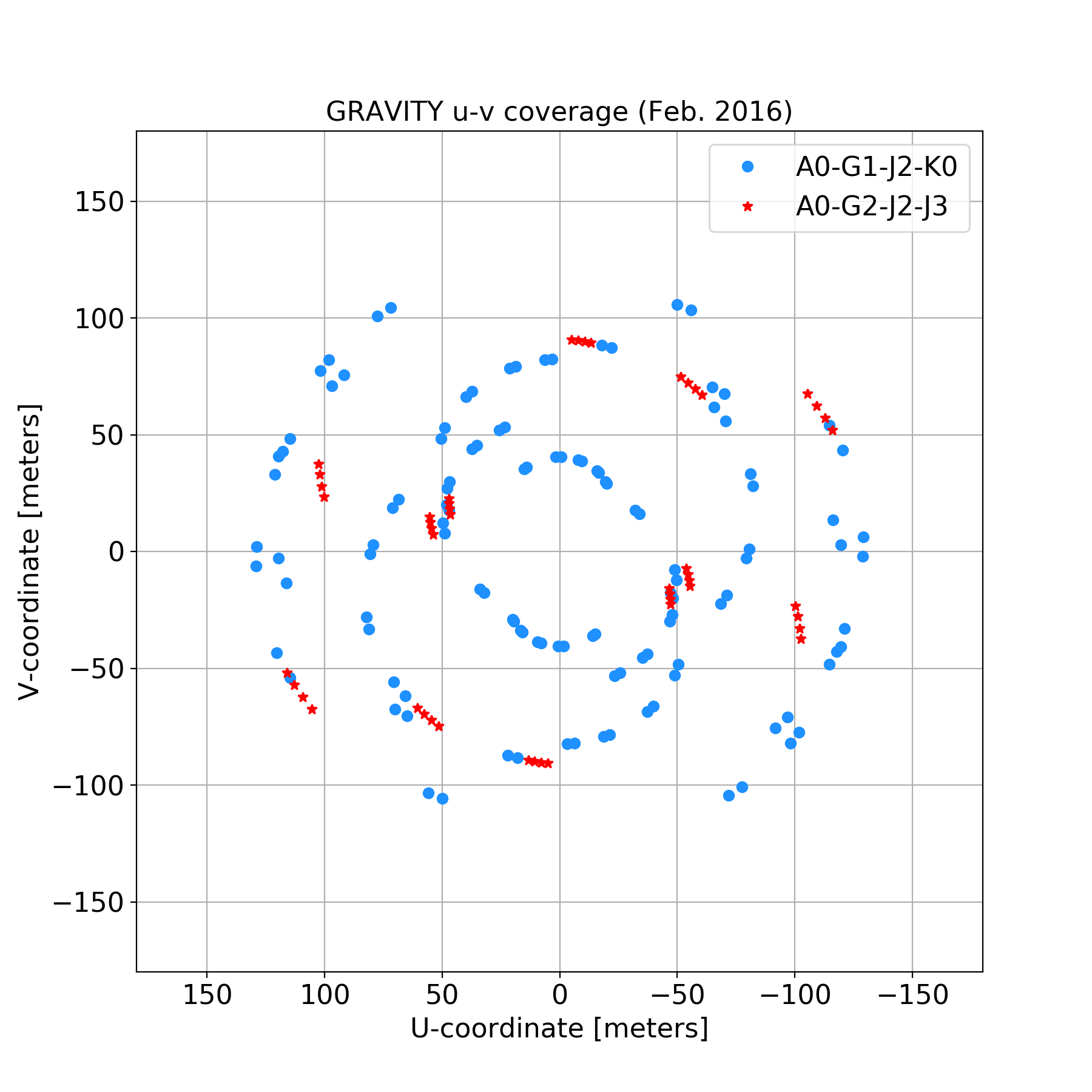}
\end{minipage}

\begin{minipage}[c]{0.4\textwidth}
\centering
    \includegraphics[width = 8 cm]{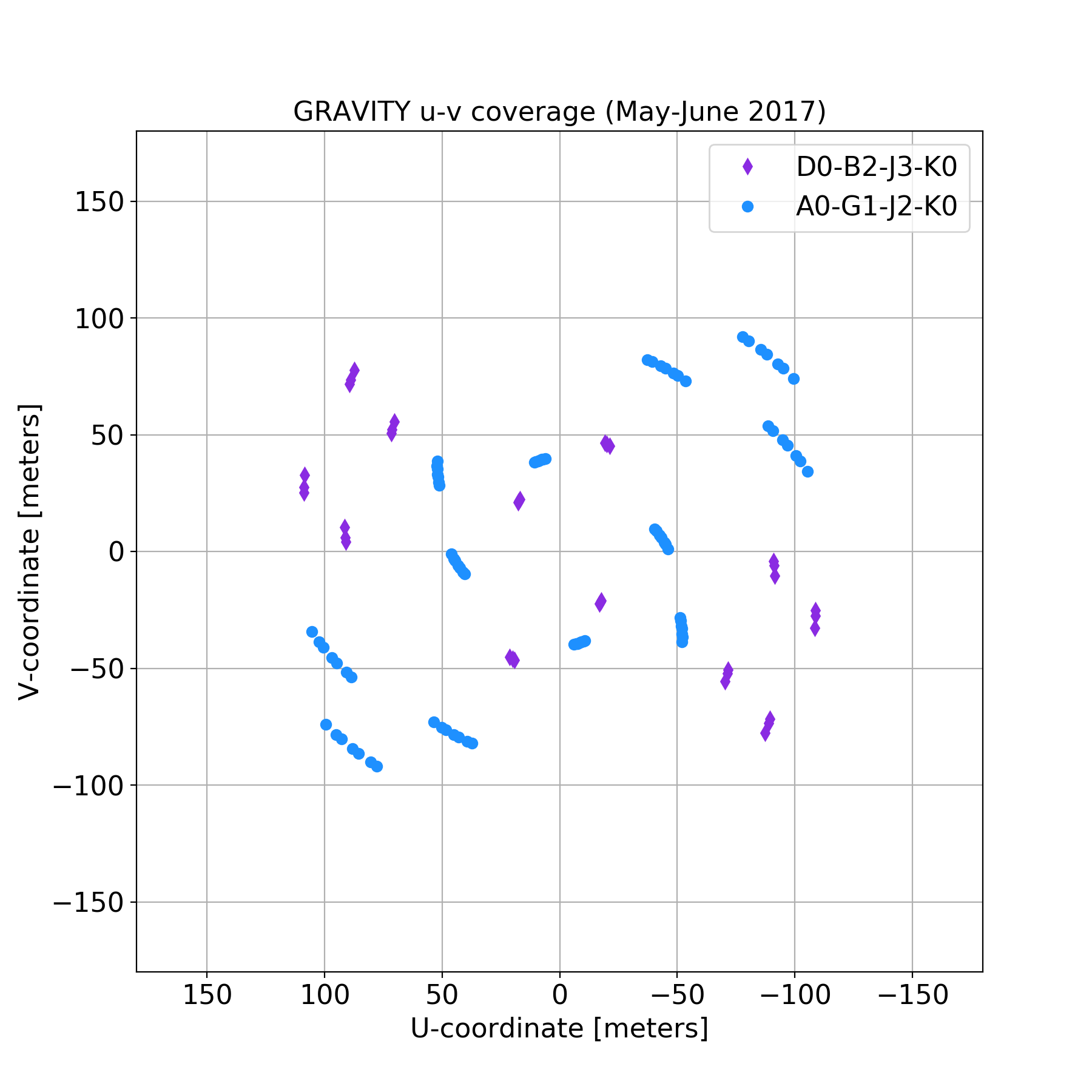}
\end{minipage}
\caption{ $\eta$ Car's $u-v$ coverages obtained during
  the GRAVITY runs in February 2016 (\textit{top}) and May-June
  2017  (\textit{bottom}). Different quadruplets are indicated with different colors.}
\label{fig:uv_plane}
\end{figure}

These aforementioned models also show that during the periastron
passage (phase $\phi
\sim$ 0.98--1.02) the acceleration zone of the 
post-shock wind of $\eta_B$ is affected by $\eta_A$. The hot wind of
$\eta_B$ pushes the primary wind outward and it ends up trapped inside
the cavity walls. The material in the walls is accelerated to
velocities larger than $\eta_A$'s wind terminal velocity, creating a layer of dense trapped primary wind. These layers have been observed as
concentric fossil wind arcs in [Fe {\sc ii}] and [Ni {\sc ii}] images at the inner 1'' obtained with the HST/STIS
camera \citep{Teodoro_2013, Gull_2016}.

To
explain the wind-wind phenomenology at 2--10 mas ($\sim$5--25 au) scales, several
attempts have been made to characterize the core of $\eta$ Car. Particularly, long-baseline infrared interferometry has been a decisive
technique for such studies. \citet{Kervella_2002} and \citet{vanBoekel_2003} resolved an
elongated optically thick region using the $K$-band (2.0--2.4 $\mu$m) beam-combiner VINCI \citep{Kervella_2000} at the Very
Large Telescope Interferometer \citep[VLTI;][]{Glindemann_2003}. Those authors
measured a size of 5--7 mas (11--15 au) for $\eta$ Car's core,
with a major to minor axis ratio $\epsilon$ = 1.25 $\pm$ 0.05, and a
PA = 134 $\pm$ 7$^{\circ}$. Using a Non-Local Thermal
Equilibrium (non-LTE) model, a
mass-loss rate of 1.6 $\pm$ 0.3 $\times$ 10$^{-3}~M_{\odot}$ yr$^{-1}$ was estimated. 

Follow-up observations \citep{Weigelt_2007} with the $K$-band beam combiner
VLTI-AMBER \citep{AMBER_Petrov_2007} using
medium (R = 1500) and high (R = 12000) spectral resolutions
of the He {\sc i} 2s-2p (2.0587 $\mu$m) and Br$\gamma$ (2.1661 $\mu$m) lines,
resolved $\eta$ Car's wind structure at angular scales as small as $\sim$ 6 mas ($\sim$
13 au). These authors measured a (50\% encircled
energy) diameter of 3.74--4.23 mas (8.6--9.7 au) for the
continuum at 2.04 $\mu$m
and 2.17 $\mu$m; a diameter of 9.6 mas (22.6 au) at the peak of the
Br$\gamma$ line; and 6.5 mas (15.3 au) at the emission peak of the He
{\sc i} 2s-2p
line. They also confirmed the presence of an elongated optically
thick continuum core with a measured axis ratio $\epsilon$ = 1.18 $\pm$ 0.10 and
a projected PA = 120$^{\circ} \pm$ 15$^{\circ}$. 

The observed non-zero differential phases
and closure phases indicate a complex extended structure across the
emission lines. To explain the Br$\gamma$-line profile and the signatures observed in the differential
and closure phases, \citet{Weigelt_2007} developed a ``rugby-ball'' model for an
optically thick, latitude dependent wind, which includes three
components: (a) a continuum spherical component; (b) a spherical
primary wind; (c) and a surrounding aspherical wind component inclined
41$^{\circ}$ from the observer's LOS. 

New high-resolution AMBER observations in 2014 allowed
\citet{Weigelt_2016} to reconstruct the first aperture-synthesis images across the Br$\gamma$
line. These images revealed an asymmetric and elongated structure,
particularly in the blue wing of the line. At velocities between $-$140
and $-$380$~\kms$, the intensity distribution of the reconstructed maps
shows a fan-shaped structure with an 8.0 mas (18.8 au) extension
to the southeast and 5.8 mas (13.6 au) to the northwest. The symmetry axis of this
elongation is at a PA = 126$^{\circ}$, which coincides with that of
the Homunculus axis. This fan-shaped morphology is consistent with the wind-wind
collision cavity
scenario described by \citet{Okazaki_2008}, \citet{Madura_2012b,
  Madura_2013}, and
  \citet{Teodoro_2013, Teodoro_2016}, with the observed emission
  originating mainly from the material flowing along the cavity in a LOS
preferential toward the southeast wall. 

Additionally to the wind-wind collision cavity, several other wind
structures were discovered in the images reported by \citet{Weigelt_2016}. At velocities between
$-$430 and $-$340$~\kms$, a
bar-like feature appears to be located southwest of the
continuum. This bar has the same PA as the more extended fossil wind structure reported by
\citet{Falcke_1996} and  \citet{Gull_2011, Gull_2016}, that may
correspond to an equatorial disk and/or toroidal material that
obscures the primary star in the LOS. At positive velocities, the
emission appears not to be as extended as in the blue-shifted part of
the line. This may be because we are looking at the back (red-shifted) part of
the primary wind that is less extended because it is not as deformed by the wind collision zone. Finally, the wind lacks any strong emission line features at velocities lower than $-$430$~\kms$ or larger than $+$400$~\kms$.           
\begin{figure*}[htp]
\centering
\includegraphics[width=\textwidth]{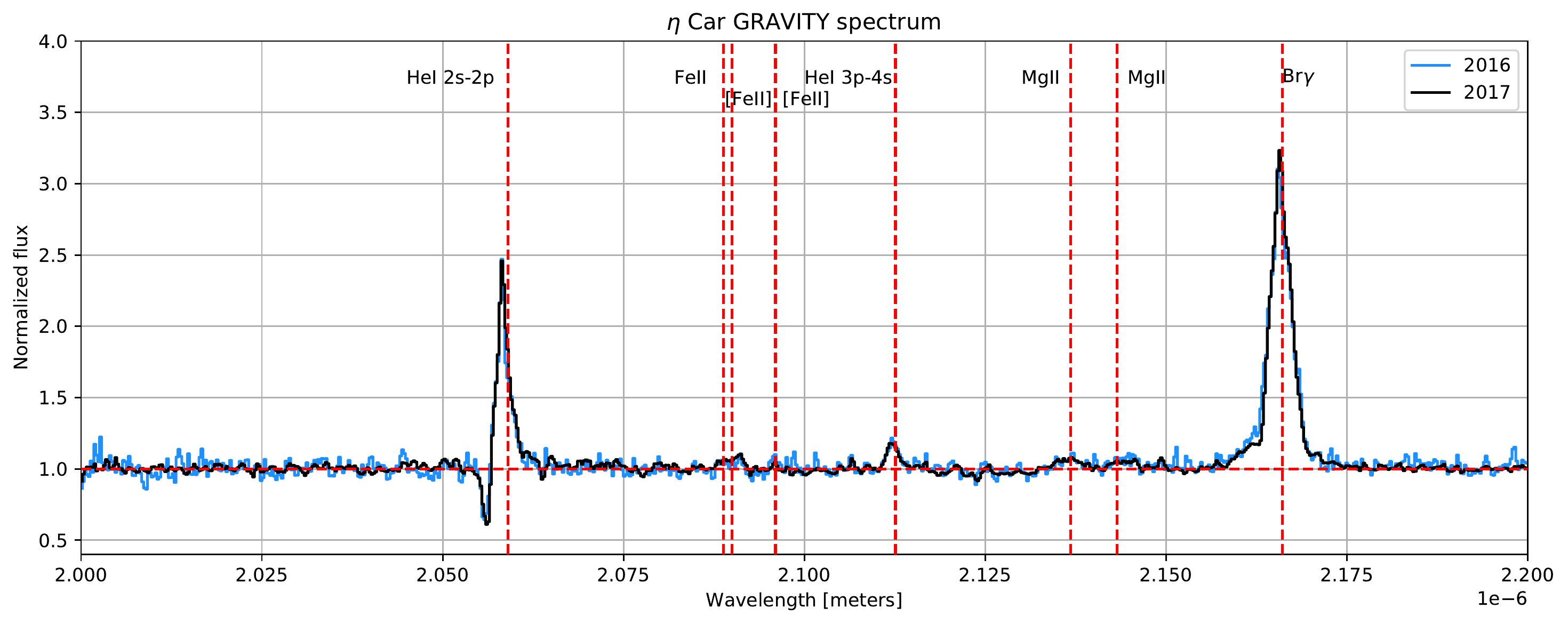}
\caption{2016 (blue straight line) and 2017 (black straight line) $\eta$ Car normalized spectra in the 2.0--2.2 $\mu$m
  bandpass. The vertical red-dashed lines indicate the spectral
  features identified in the spectrum. }
\label{fig:GRAVITY_spectrum} 
\end{figure*}

This work presents VLTI-GRAVITY chromatic imaging of $\eta$ Car's core across two spectral lines in the infrared
$K$-band: He {\sc i} 2s-2p and
Br$\gamma$. The paper is outlined as follows:  Sect.\,\ref{sec:data_reduction}
presents our GRAVITY observations and data reduction. In
Sect.\,\ref{sec:results} the analyses of the interferometric
observables, and the details of the imaging procedure are described. In Sect.\,\ref{sec:discussion} our results are discussed and,
finally, in Sect.\,\ref{sec:conclusions} the conclusions are presented. 

\section{Observations and Data reduction \label{sec:data_reduction}}

\subsection{Observations}

The milliarcsecond resolution of GRAVITY \citep{Eisenhauer_2008, Eisenhauer_2011, Eisenhauer_2017}  enables spectrally resolved
interferometric imaging of the central wind region of $\eta$ Car. At an apparent magnitude of m$_K$ = 0.94 mag, the target is bright enough for observations with the 1.8-meter Auxiliary
Telescopes (ATs). $\eta$ Car was observed during the nights of
February 24th and 27th, 2016, as part of the commissioning phase of the
instrument, and during the nights of May 30th and July 1st 2017, as part of the GRAVITY Guaranteed Time Observations (GTO). The
observations were carried out using the highest spectral resolution
(R = 4000) of the GRAVITY beam combiner, together with the split-polarization and
single-field modes of the instrument. With this configuration, GRAVITY
splits the incoming light of the science target equally between the fringe
tracker and science beam combiner to simultaneously produce
interference fringes in both. While the
science beam combiner disperses the light at the desired spectral
resolution, the fringe tracker works with a low-spectral resolution of R$\sim$22
\citep{Gillessen_2010} but at a high temporal sampling ($\sim$1
kHz). This allows to correct for the atmospheric piston, and to
stabilize the fringes of the science
beam combiner. 

For the 2016 observations, ten data sets were recorded with the
A0-G1-J2-K0 array plus four more with the A0-G2-J2-J3
configuration. The $u-v$ coverage obtained (Fig.\,\ref{fig:uv_plane}) provides a maximum projected baseline of
$\sim$130 m (J2-A0) that corresponds to a
maximum angular resolution ($\theta$=$\lambda$/2$B$; where B is the
maximum baseline) of $\theta$ = 1.75 mas, at a
central wavelength of $\lambda_0$= 2.2 $\mu$m. The
2017 observations comprise three data sets with the
D0-B2-J3-K0 array plus seven data sets with the A0-G1-J2-K0 array. For
this second epoch the
maximum baseline length (J2-A0) was around 122 m ($\theta$ = 1.85 mas). However, since most of the longest baselines for both
imaging epochs are of 100 m, we adopted a mean maximum angular
resolution $\theta$ = 2.26 mas for our imaging program. Tables
\ref{tab:2016_obs} and \ref{tab:2017_obs} list individual data sets
and observing conditions. 

\subsection{Data reduction}

The interferometric observables (squared visibilities, closure phases,
and differential phases) as well as the source's spectrum
were obtained using versions 0.9.0 and 1.0.7 of the GRAVITY data reduction
software\footnote{http://www.eso.org/sci/software/pipelines/gravity/gravity-pipe-recipes.html}
\citep{Lapeyrere_2014}. More details on the data reduction procedure are provided in \citet{Sanchez-Bermudez_2017}. To calibrate the interferometric observables, interleaved observations
of the science target and a point-like source were
performed. We used the K3
II star HD\,89682 ($K$-band Uniform Disk diameter d$_{UD}$ = 2.88 mas) as
interferometric calibrator for both epochs. Before analyzing the data, all the
squared visibility (V$^2$) points with a signal-to-noise ratio (S/N) $\leq$ 5 and
closure phases with $\sigma_{cp} \geq$ 40$^{\circ}$ were excluded from
the analysis. 

\begin{figure*}[htp]
\centering
\includegraphics[width=14 cm]{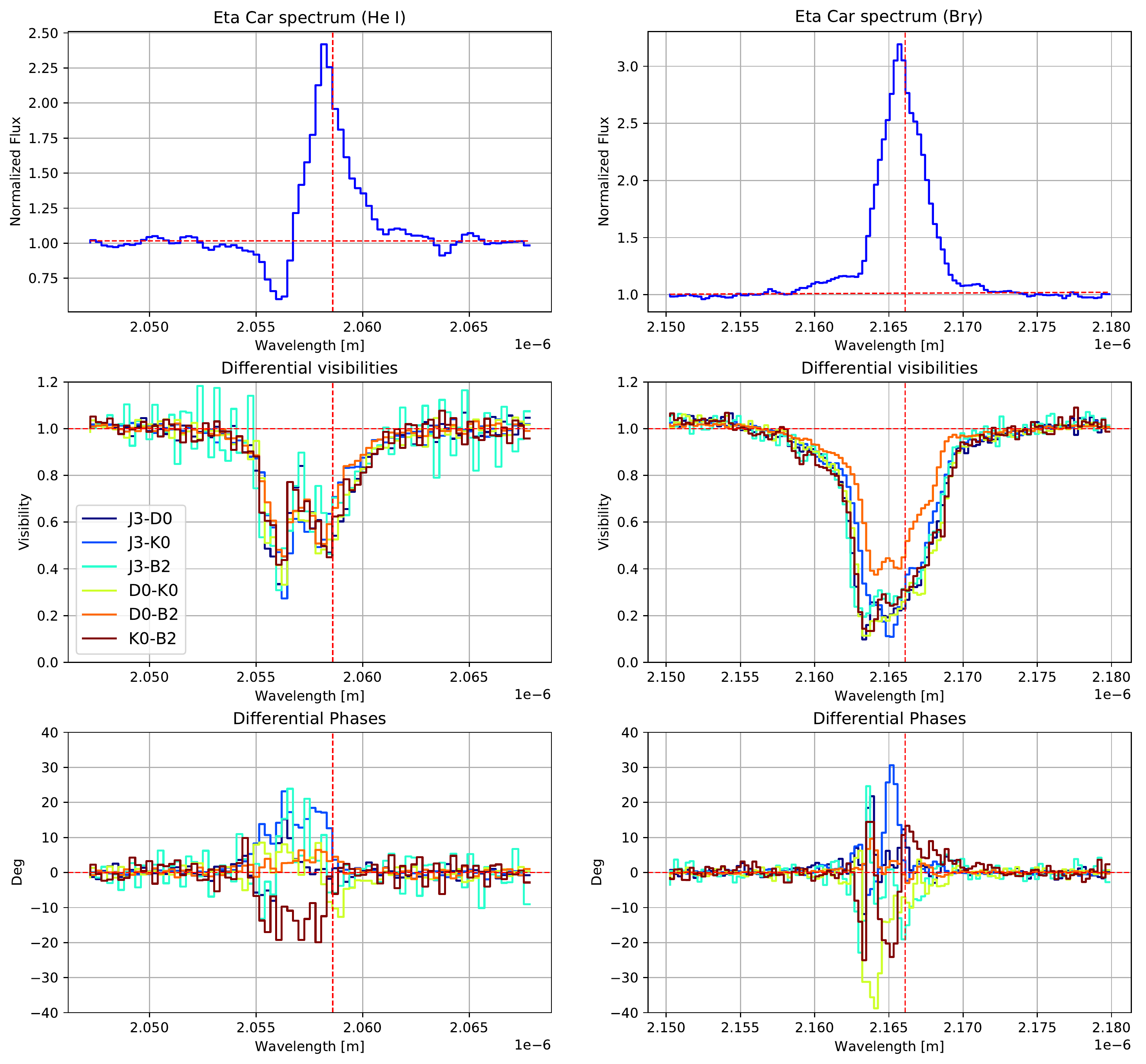}
\caption{Normalized GRAVITY spectrum at the position of the He {\sc i} 2s-2p
  and Br$\gamma$ lines. The calibrated differential visibilities
  (\textit{middle}) and phases (\textit{bottom}) of one of
  the GRAVITY data sets (MJD: 57903.9840) are also shown. In the
  \textit{middle} and \textit{bottom} panels, different colors
  correspond to each one of the baselines in the data set (see label
  on the plots). The
  dashed-red vertical lines indicate the reference wavelength of the
  emission feature, and the dashed-horizontal lines show the reference value of the
  continuum baseline. }
\label{fig:GRAVITY_observables} 
\end{figure*}

\subsection{Calibration of $\eta$ Car's integrated spectrum \label{sec:spectrum}}

For each one of the data sets, four samples of $\eta$ Car's integrated spectrum were obtained with
GRAVITY. The data reduction software delivers the spectrum flattened
by the instrumental transfer function obtained from the Pixel to
Visibility Matrix \citep[P2VM;][]{AMBER_Petrov_2007}. Additionally, GRAVITY uses the
internal calibration unit  \citep{Blind_2014} to obtain the fiber
wavelength scale, producing a wavelength map with a precision
of $\Delta_{\lambda}$ = 2 nm. Apart from this preliminary
calibrations, we
had to correct the $\eta$ Car spectrum for the atmospheric transfer
function together with a more precise wavelength calibration. For this
purpose, we used the following method:

\begin{enumerate}
\item  A weighted average science and calibrator spectra
were computed from the different samples obtained with GRAVITY.

\item To refine the wavelength calibration, we used a series of
  telluric lines across the spectrum of our interferometric
  calibrator. A Gaussian was fitted to each one of the lines profiles
 recording their peak positions. The high-resolution (R = 40000)
 telluric spectrum at the Kitt Peak observatory was then used
 as reference to re-calibrate the GRAVITY wavelength map. For this, we first
 degraded the reference spectrum to the GRAVITY resolution and then,
 we identified a mean wavelength shift of the calibrator's tellurics
 from the Kitt Peak ones. Finally, the GRAVITY wavelength map was
 corrected. With this method, we reached a 1$\sigma$ wavelength
 calibration error relative to the reference spectrum of 1.23 \AA\, ($\Delta \nu$
 = 16.8 $\kms$ at $\lambda_0$ = 2.2 $\mu$m) for the 2016
 data, and 0.67 \AA\, ($\Delta \nu$ = 9.1 $\kms$) for the 2017
 data, respectively. A similar calibration method was used by
 \citet{Weigelt_2007, Weigelt_2016} on the  previously reported $\eta$
 Car AMBER data .

\item Once the wavelength calibration was performed, we remove the
  atmospheric transfer function from the $\eta$ Car spectrum by obtaining the ratio of the
  science target and of the interferometric calibrator spectra. Since the interferometric
calibrator, HD\,89682, is a K3 II star, we have to correct for its intrinsic
photospheric lines (e.g., the CO bands seen in absorption from 2.29 $\mu$m onward). For this purpose, we used a theoretical BT-Settl\footnote{http://svo2.cab.inta-csic.es/theory/newov2/}
model \citep[see ][]{Allard_2011} of
a star with a T$_{\mathrm{eff}}$ = 4200 $K$, a log(g) = 1.5, and Z=
$-$0.5. Figure\,\ref{fig:GRAVITY_spectrum} displays the normalized $\eta$ Car
spectrum with all the lines, between 2.0--2.2 $\mu$m, with interferometric signals different
from the continuum labeled. Although the 2016 spectrum appears to be
slightly noisier than the 2017 one, no significant differences were found between the
two GRAVITY epochs. This work focuses on the interferometric imaging
of $\eta$ Car across He {\sc i} 2s-2p and Br$\gamma$. Figure\,\ref{fig:GRAVITY_observables} displays the normalized $\eta$ Car
spectrum at these wavelengths, together with
the differential visibilities and phases of one of our
interferometric data sets. Notice the strong changes of the observables between the
continuum and the lines. 

\end{enumerate}
     
\subsection{Ancillary FEROS spectrum}

\begin{figure*}[htp]
\centering
\includegraphics[width=12 cm]{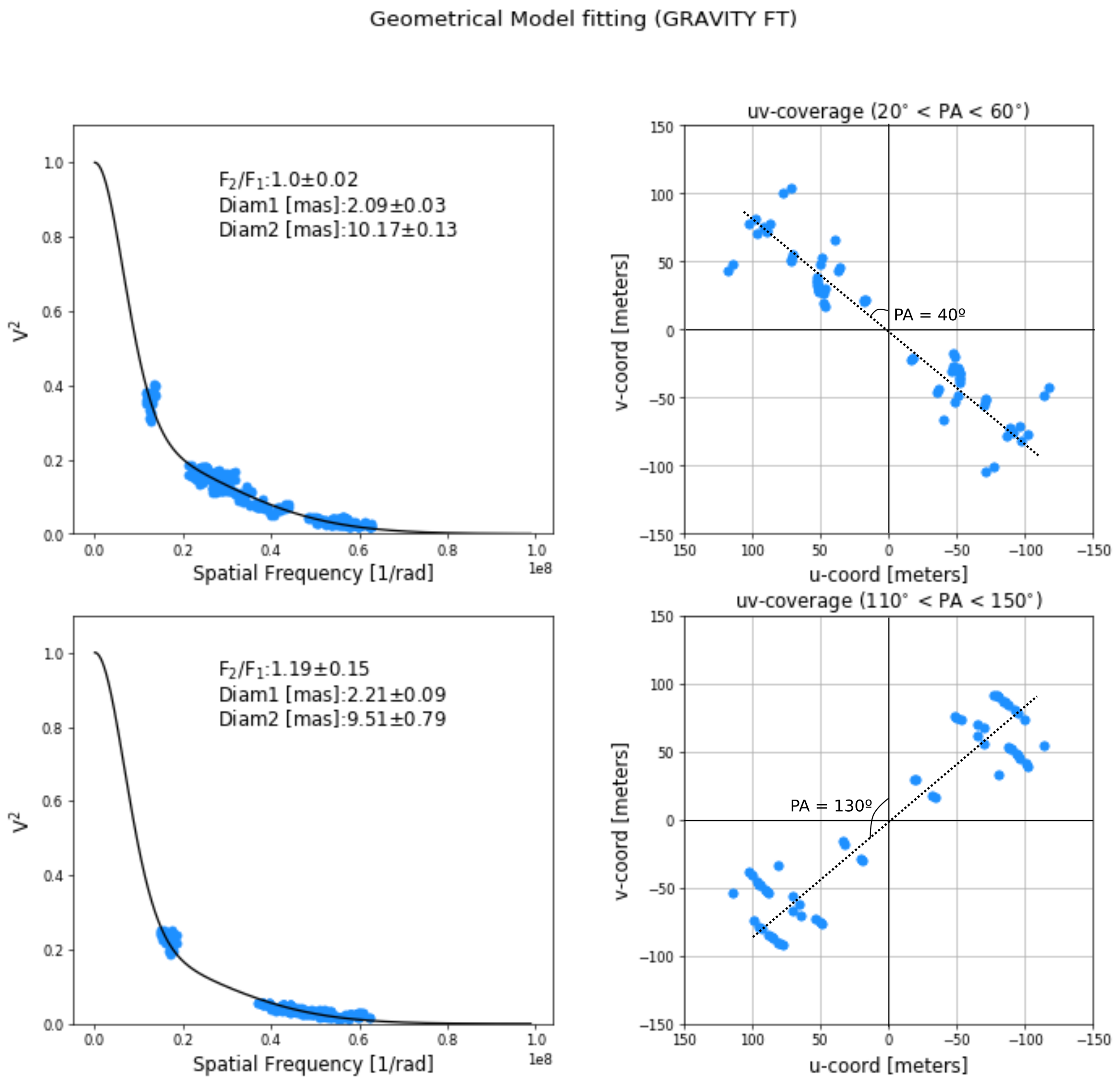}
\caption{The \textit{left} panels show the best-fit geometrical model to the Fringe Tracker $V^2$
  data. The model is plotted with a black line, the data
  are shown with blue dots and the model parameters (with their
  uncertainties) are indicated on each panel. The \textit{right} panels display the
  visibility points used for each model fitting at two different
  position angles,   PA$_{\bot}$ = 40$^{\circ} \pm$
20$^{\circ}$ and PA$_{\parallel}$ =  130$^{\circ} \pm$
20$^{\circ}$. }
\label{fig:GRAVITY_geom} 
\end{figure*}

 To complement the analysis of the $\eta$ Car GRAVITY spectrum, we
 obtained four optical high-resolution spectra of the source using the
Fiber-fed Extended Range Optical Spectrograph \citep[FEROS;][]{Kaufer_1998} at the MPG 2.2m telescope at La Silla
Observatory in Chile. FEROS covers the entire optical spectral range
from 3600\, \AA\, to 9200\, \AA\, and provides a spectral resolution of
R = 48\,000. All spectra were obtained using the object-sky mode
where one of the two fibers is positioned on the target star and the
other fiber is simultaneously fed with the sky background. The exposure
time was 15\,s and 30\,s for every two spectra. The reduction and
calibration of the raw data were performed using the CERES pipeline
\citep{Brahm_2017}. Figure \ref{fig:FEROS_spectrum} displays a normalized mean
spectrum of the four samples.

\section{Data Analysis \label{sec:results}}

\begin{figure*}[htp]
\centering
\includegraphics[width=19 cm]{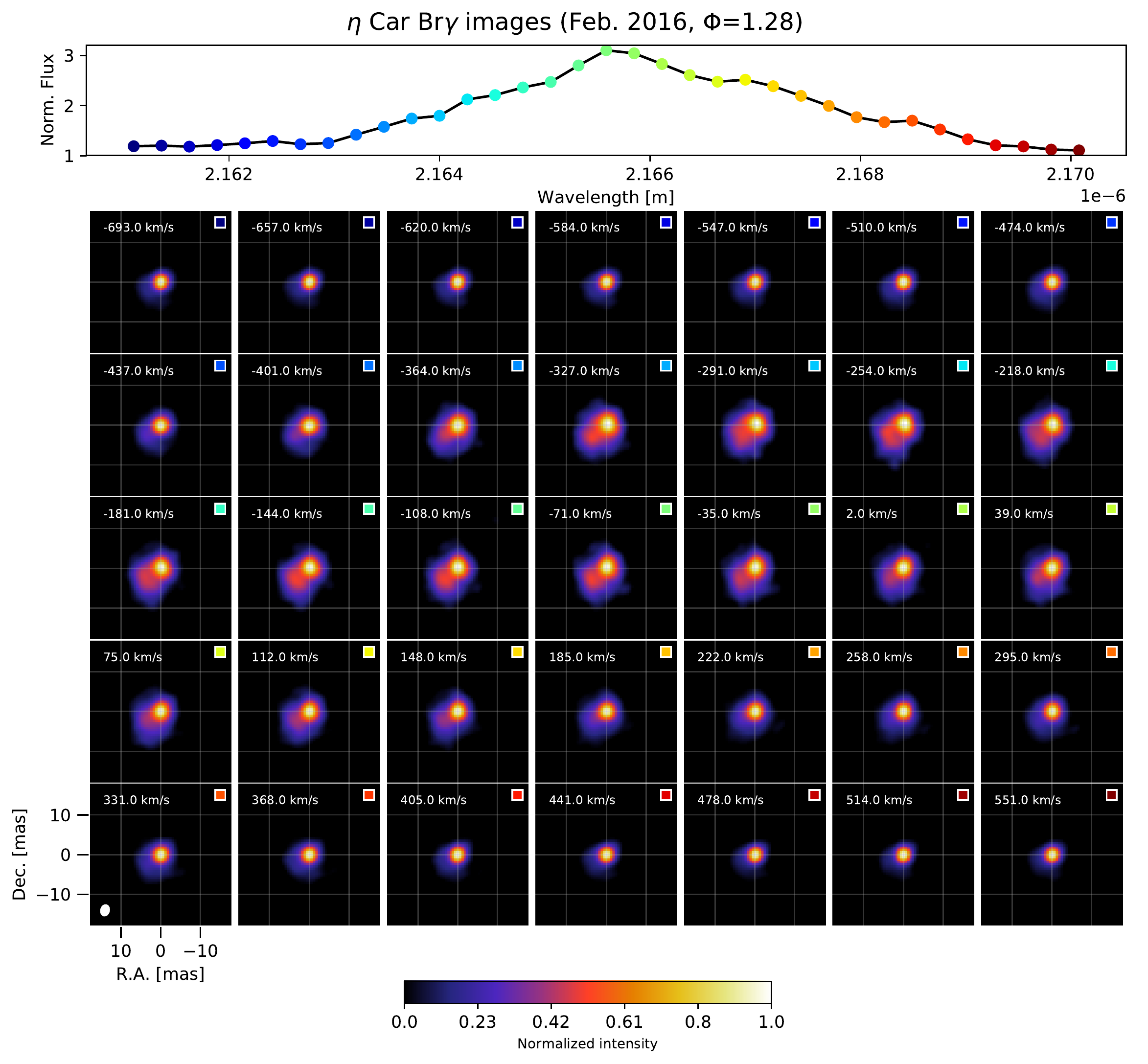}
\caption{Br$\gamma$ interferometric aperture synthesis images from the
  Feb. 2016 data. The
  Doppler velocity of each frame is labeled in the images. For all the panels, east is to the left and north to the
  top. The displayed FOV corresponds to 36$\times$36
  mas. The small white ellipse shown in the lowermost-left panel corresponds
  to the synthesized primary beam (the detailed PSF is shown in Fig.\,\ref{fig:BrG_p1_june2017}). Above all the images, the GRAVITY
  spectrum is shown and the different positions where the images are
  reconstructed across the line are labeled with a colored square,
  which is also plotted in the images for an easy identification.}
\label{fig:BrG_p1_feb2016} 
\end{figure*}

\begin{figure*}[htp]
\centering
\includegraphics[width=19 cm]{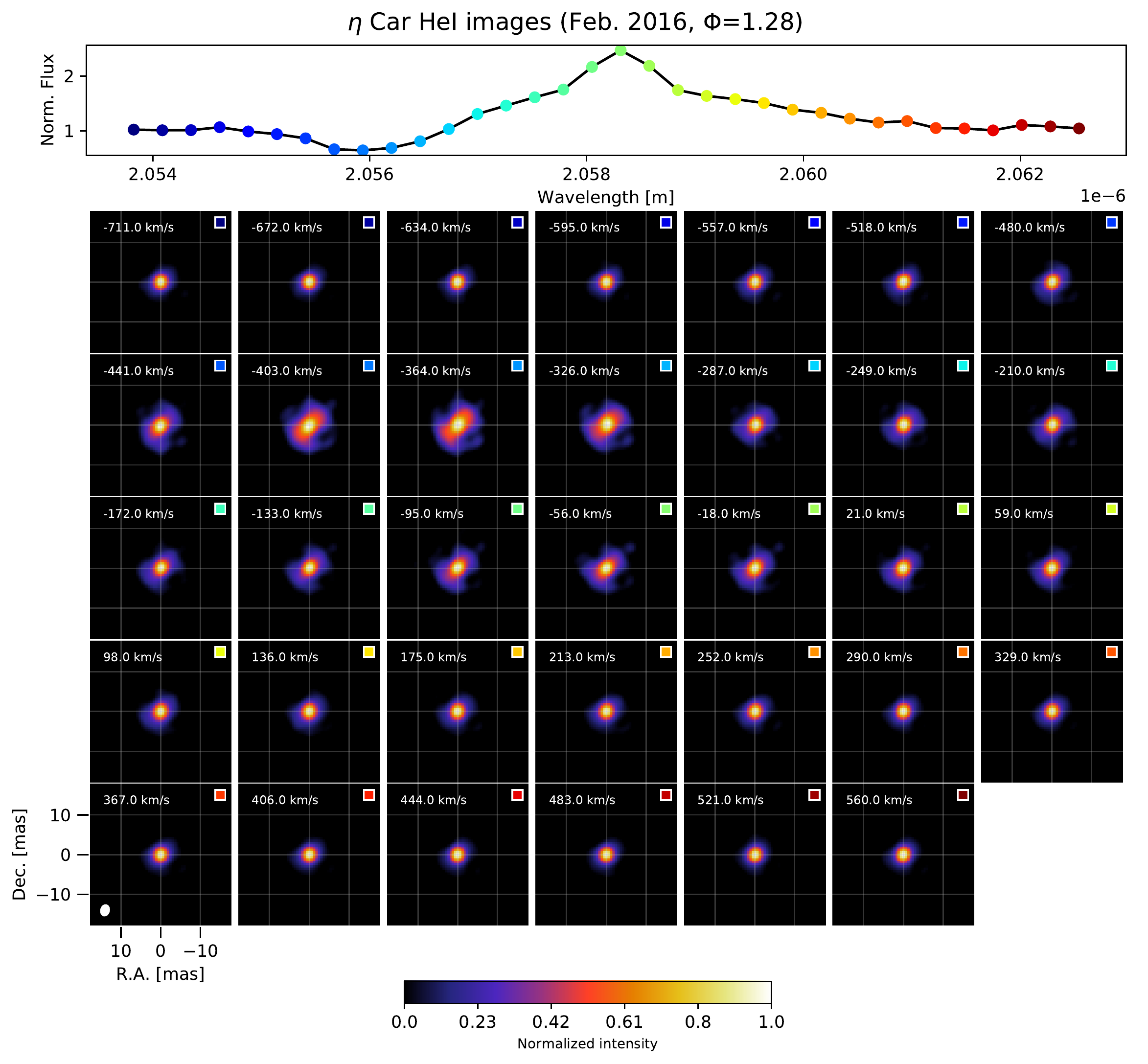}
\caption{He {\sc i} interferometric aperture synthesis images from the Feb. 2016
  data. The maps are as described in Figure \ref{fig:BrG_p1_feb2016}.}
\label{fig:HeI_p1_feb2016} 
\end{figure*}

\subsection{Geometric model to the GRAVITY Fringe Tracker data
\label{sec:geom_model}}

To obtain an angular measurement of the continuum size, a geometrical
model was applied to the $V^2$ Fringe Tracker data. The model consists of
two Gaussian disks with different angular sizes tracing the compact and
extended brightness distribution of the source. The expression used to
estimate the complex visibilities, V($u,\,v$), is the following one:  

\begin{equation}
\mathrm{V}(u,v) = \frac{\mathrm{G}(\theta_1, u,
  v)+F_{\mathrm{2}}/F_{\mathrm{1}}\mathrm{G}(\theta_2, u, v)}{1 + F_{\mathrm{2}}/F_{\mathrm{1}}}\,
\end{equation}

with 

\begin{equation}
\mathrm{G}(\theta, u, v) = \mathrm{exp}\left(\frac{-\pi \theta \sqrt{(u^{\mathrm{2}} + v^{\mathrm{2}})}}{4\mathrm{ln(2)}}\right)\,,
\end{equation}

where $F_{\mathrm{2}}/F_{\mathrm{1}}$ is the flux ratio between the two
Gaussian disks,  ($u$, $v$) are components of the spatial
frequency sampled per each observed visibility point
($u$=$B_{\mathrm{x}}$/$\lambda$ and $v$=$B_{\mathrm{y}}$/$\lambda$), and $\theta$ is the fitted
full-width-at-half-maximum (FWHM) angular
size for each one of the components. 

Since previous studies suggest an elongation of $\eta$ Car's core
along the projected PA of the Homunculus semi-major
axis, we applied our geometrical model to $V^2$ at two orientations. To cover
PAs perpendicular and parallel to the semi-major axis of the nebula, we
used all the $V^2$ points at PA$_{\bot}$ = 40$^{\circ} \pm$
20$^{\circ}$ and at PA$_{\parallel}$ =  130$^{\circ} \pm$
20$^{\circ}$, respectively. The model optimization was done using a dedicated Markov-Chain Monte-Carlo (MCMC) routine based on
the python software \texttt{pymc} \citep{pymc_2010}. To account
for the standard deviations and correlations of the model parameters,
we explore the solution space over 10000 different models, considering
the first 5000 iterations as part of the burn-in phase and accepting
only one every three draws of the Monte Carlo. Initial linear distributions
were assumed for all the parameters. To avoid an underestimation of the error-bars,
the waist of the posterior Gaussian likelihood
distribution was also marginalized for each one of the fitted
parameters. Figure \ref{fig:GRAVITY_geom}
shows the best-fit model and its
uncertainties. Table\,\ref{tab:best_fit_model} displays the best-fit
parameters of our geometrical model and Figure \ref{fig:post_dist} shows the 2D posterior distributions of the fitted parameters. 

\begin{table}
\caption{Best-fit parameters of the geometrical model}
\label{tab:best_fit_model}
\centering
\begin{tabular}{l l l l} 
\hline\hline
PA & Parameter & Value & 1-$\sigma$ \\
\hline
40$^{\circ}\,\pm$ 20$^{\circ}$ & $\theta_1$ [mas]& 2.09 & 0.03\\
&$\theta_2$ [mas] & 10.17 & 0.13 \\
&$F_2$ / $F_1$ & 1.00 & 0.02 \\
\hline
130$^{\circ}\,\pm$ 20$^{\circ}$ & $\theta_1$ [mas] & 2.21 & 0.09\\
&$\theta_2$ [mas] & 9.51 & 0.79\\
&$F_2$ / $F_1$ & 1.19 & 0.15\\
\hline
\end{tabular}
\end{table}

\subsection{Image reconstruction of the Science Beam Combiner data} 

\begin{figure*}[htp]
\centering
\includegraphics[width=14 cm]{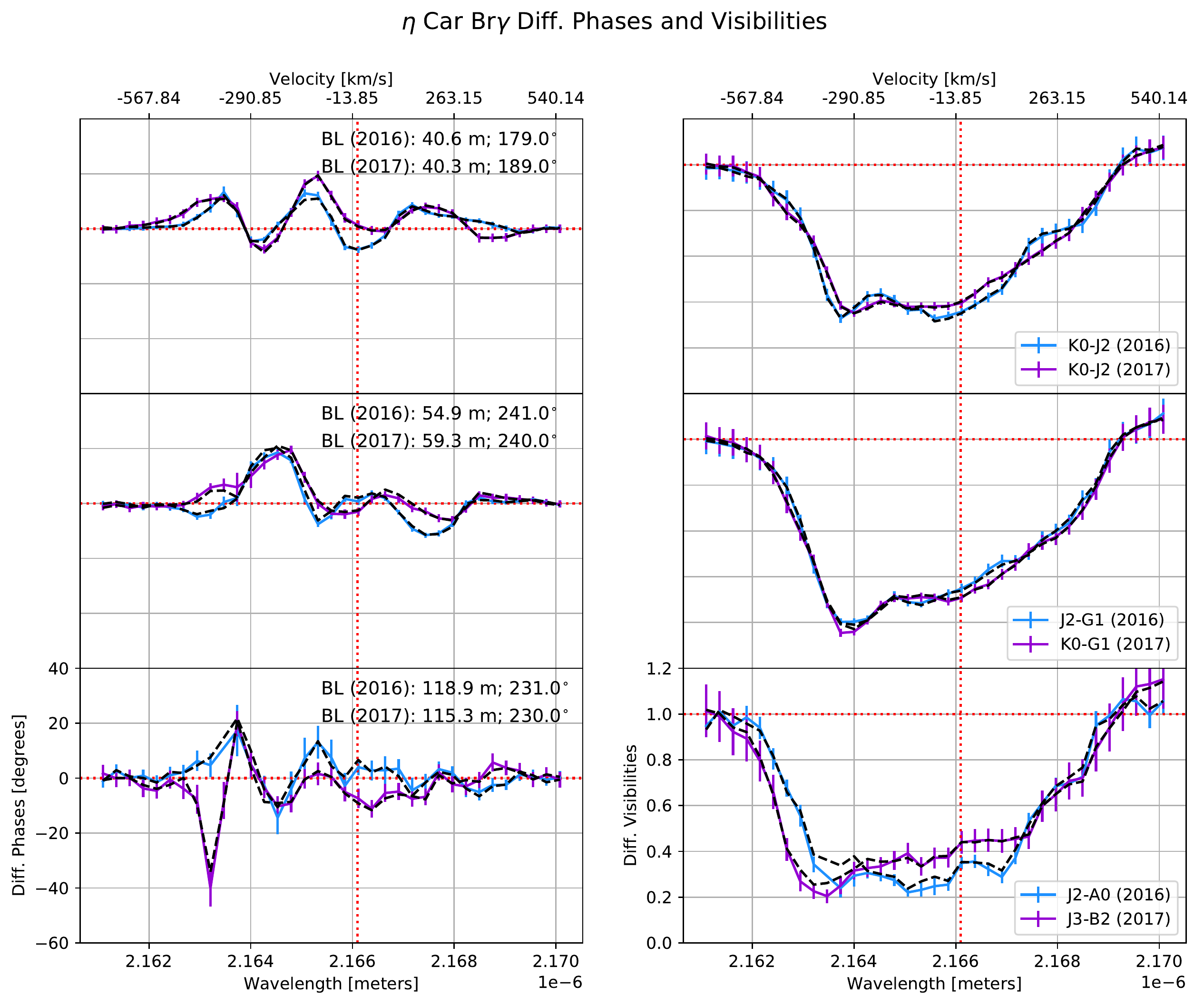}
\caption{Differential phases (\textit{left}) and
  differential visibilities (\textit{right}) between the
  continuum and the line emission for the 2016 (blue line) and 2017 data
  (purple line). The (reference) continuum phase and visibility were
  estimated using the first and last five channels in the imaged bandpass. Each one of the rows corresponds to a
  similar baseline in both epochs, tracing large (top), intermediate
  (middle) and small (bottom) spatial scales. The baseline lengths and position
  angles are labeled on the left panels, while the baseline stations are
  labeled on the right ones. The black-dashed lines correspond to the
  differential quantities extracted from the unconvolved reconstructed images. The
  vertical red-dotted line marks the systemic velocity of Br$\gamma$. The horizontal red-dotted lines show the
  continuum baselines.}
\label{fig:diffphases_comparison} 
\end{figure*}

\subsubsection{Chromatic reconstruction: Br$\gamma$ and He {\sc i}}

The high accuracy of the interferometric observables allows a chromatic
image reconstruction of $\eta$ Car across the Br$\gamma$  and He\,{\sc
i} 2s-2p lines. For this purpose, we used \texttt{SQUEEZE}\footnote{https://github.com/fabienbaron/squeeze} \citep{Baron_2010}, an
interferometry imaging software that allows simultaneous fitting of
the visibility amplitudes,
squared visibilities, closure phases, and chromatic differential phases (or
combinations of them). At infrared
wavelengths, image reconstruction from interferometric data is
constrained mainly by the (i) sparse baseline coverage and (ii) the
lack of absolute phase information. Therefore, \texttt{SQUEEZE} applies a regularized
minimization of the form: 

\begin{equation}
\boldsymbol{x}_{\mathrm{ML}}=\underset{\boldsymbol{x}}{\textrm{argmin}}[1/2 \chi^2(\boldsymbol{x})+\sum_{i}^{n}\mu_i R(\boldsymbol{x})_i]\,,\\
\end{equation}

where $\chi^2(\boldsymbol{x})$ is the likelihood of our data to a
given imaging model,  $R(\boldsymbol{x})_i$ are the (prior) regularization functions
used and $\mu_i$ the weighting factors that trade-off
between the likelihood and priors. For $\eta$ Car, we used a combination of the following three different
\texttt{SQUEEZE} regularizers:

\begin{enumerate}
\item To avoid spurious point-like sources in the field-of-view (FOV), we applied a spatial
  L0-norm.

\item To enhance the extended structure expected
  in some of the mapped spectral channels, a Laplacian was used. 

\item To ensure
  spectral continuity all over the mapped emission lines, a spectral
  L2-norm was applied.
\end{enumerate}

For the minimization, \texttt{SQUEEZE} uses a Simulated Annealing
Monte-Carlo algorithm as the engine for the reconstruction. We created 15 chains with 250 iterations each to find the most
probable image that best-fit our interferometric data. We used a
67$\times$67 pixel grid with a scale of 0.6 mas/pixel. As initial
point for the reconstruction, we chose a Gaussian
with a FWHM of 2 mas centered in the pixel grid, and containing 50\%
of the total flux. We
simultaneously fitted the $V^2$, closure phases, and wavelength-differential
phases. The differential phase is defined in \texttt{SQUEEZE} of the
following form:

\begin{equation}
\phi(\lambda_i) = \phi_0(\lambda_i) - \phi_{\mathrm{ref}}.\,\\
\end{equation}

where $\phi_0(\lambda_i)$ is the measured phase at channel $i$, and
$\phi_{\mathrm{ref}}$ is the reference phase. The \texttt{SQUEEZE}
implementation of $\phi (\lambda_i)$ uses the \texttt{REFMAP} table, of
the \texttt{OI\_VIS} extension in the \texttt{OIFITS v2} files
\citep{Pauls_2005, Duvert_2015}, to
determine which channels are used to define the reference phase
$\phi_{\mathrm{ref}}$. In our case, we defined $\phi_{\mathrm{ref}}$ as the average of the
measured phases over the full spectral bandpass used for the reconstruction, with exception of the working channel itself, in the following way:

\begin{equation}
\phi_{\mathrm{ref}} =\langle \phi_0(\lambda_k) \rangle_{\lambda_k \not= \lambda_i} \\
\end{equation}

With this definition, the reference phase varies from channel to
channel but avoids quadratic bias terms \citep[a similar
definition of $\phi(\lambda_i)$ has been used before to analyze AMBER
data; see e.g., ][]{Millour_2006, Millour_2007, AMBER_Petrov_2007}. While the $V^2$ data
measure the flux contribution at different spatial scales and the
closure phases the asymmetry of the source, the
wavelength-differential phases give information about the relative
flux-centroid displacement between a given wavelength and the
reference one.

For Br$\gamma$, 35 channels between 2.160 $\mu$m and
2.171 $\mu$m were imaged, while for He {\sc i}, 34 channels between
2.054 $\mu$m and 2.063 $\mu$m were used. The final mean images were created through the
following procedure:

\begin{enumerate}
\item We ranked the converged chains, assigning the highest score to the
  one with the global $\chi^2$ closest to unity. From the ranked chains, we
  select the best five of them to create the final
  images. 


\item To align the selected cubes of images to a common reference pixel
  position, a mean centroid position of the continuum was
  estimated. For this purpose, the first and last five channels of
  each cube of images were used. The individual centroid positions were computed using a
  mask of 5 pixels around the maxima of the images at every channel used as
  continuum reference. Then, their mean value was computed. All the images in the cube were shifted, with sub-pixel accuracy ($\sim$ 0.1 pixels), from this mean centroid position to the
  center of the image defined at position [34,34] in the pixel grid.

\item From the aligned cubes, we compute a mean image per
  wavelength. Finally, since most of the spatial frequencies in the $u
  - v$ coverages correspond to 100 m baselines, each image was smoothed with a 2D Gaussian
  with a FWHM equivalent to $\lambda_0$ /
  2$B_{\mathrm{max}}$ = 2.26 mas. 

\end{enumerate}

For both lines, reconstruction tests with the s- and p-polarization
data were done independently, however, no significant morphological
changes were observed.  The images here presented correspond to the
p-polarization data. Figures
\ref{fig:BrG_p1_feb2016} and \ref{fig:HeI_p1_feb2016} show the
mean-reconstructed 2016 Br$\gamma$ and He {\sc i} images,
respectively. The 2017 images are shown in Figures
\ref{fig:BrG_p1_june2017} and \ref{fig:HeI_p1_june2017} in the
Appendix. To inspect the quality of the reconstruction,
Figures between \ref{fig:BrG_p1_v2_feb2016} and
\ref{fig:HeI_p1_cp_june2017} in the Appendix \ref{sec:fit_v2_cp}
display the $V^2$ and closure phases from the recovered images
together with the data. 

\subsubsection{Effects of the $u-v$ coverage on the reconstructed
  images \label{sec:uv_effects}}

Due to the interacting winds of the eccentric binary at the core of $\eta$ Car, we expect changes in
the spatial distribution of the flux at different orbital phases. This condition could explain
the differences in the observed morphologies between our two imaging
epochs. However, while the $u-v$ coverage
of the 2016 data is homogeneous, the 2017 $u-v$ coverage is more
sparse. The sparseness of the 2017 $u-v$ coverage was caused,
partially, because the target was observable above 35$^{\circ}$ over
the horizon only
for a few hours at the beginning of the night. 

Therefore, the
long baselines of our interferometric array could not properly sample north-south orientations of the $u-v$ space. This sparseness in the
$u-v$ coverage produced a clumpy fine structure and a cross-like shape, superimposed on the source's
emission, on the reconstructed images. This cross-like structure is caused by the strong secondary
lobes and the elongation of the primary beam (see
Fig.\,\ref{fig:dirty_beam}). This effect could
mimic or mask real changes in the morphology associated with the physics of the
wind-wind interactions with the synthesized beam and/or artifacts
in the reconstruction process. 

To detect the morphological changes between two epochs
associated to the physical conditions of the source and not caused by
the $u-v$ coverage, we searched for coincidental baselines (in size and
position angle) at both epochs, to compare the differential phases
and visibilities between them. This test allows us to monitor the flux
centroid evolution between the continuum and the lines together with the changes observed in the flux distribution. Three baselines, tracing the
small, intermediate and large angular scales were
selected. Additionally, we compute the differential
observables from the reconstructed images to monitor the goodness of
the images recovering the possible morphological changes.

Figure\,\ref{fig:diffphases_comparison}
displays, as an example, the
results of this comparison for Br$\gamma$. For the short ($\theta =$ 5.7
mas; 13.4 au) and  intermediate ($\theta =$ 3.8 mas; 8.4 au)
baselines, the overall signatures of the
centroid and of the flux distribution are similar at both epochs, with
more prominent centroid changes at blue-shifted velocities than at the red-shifted side. The long ($\theta =$1.9
mas; 4.46 au) baseline shows more drastic differences at blue-shifted
velocities between the two epochs. For example, while the differential phases of
the 2017 data exhibit
an upside-down double-peak profile at velocities between $-$420 $\kms$ and
$-$170 $\kms$, the
2016 shows a single and less prominent peak. 

These changes in the observables for coincidental baselines provide
evidence of modifications in the brightness distribution between the
two epochs. However, despite the good fit of the 2017 images to the
observables (see Figs.\,\ref{fig:BrG_p1_v2_june2017}, \ref{fig:BrG_p1_cp_june2017}, \ref{fig:HeI_p1_v2_june2017},
and \ref{fig:HeI_p1_cp_june2017}), it is clear that the sparse $u-v$ coverage of
the second epoch strongly detriments the quality of the reconstructed
images. Therefore, they cannot be used for a proper
(direct) comparison of the global morphology observed in the 2016
data. Thus, the 2017 images were excluded for the following analysis
and discussion. 

\section{Discussion \label{sec:discussion}}

\begin{figure}[htp]
\centering
\includegraphics[width=\columnwidth]{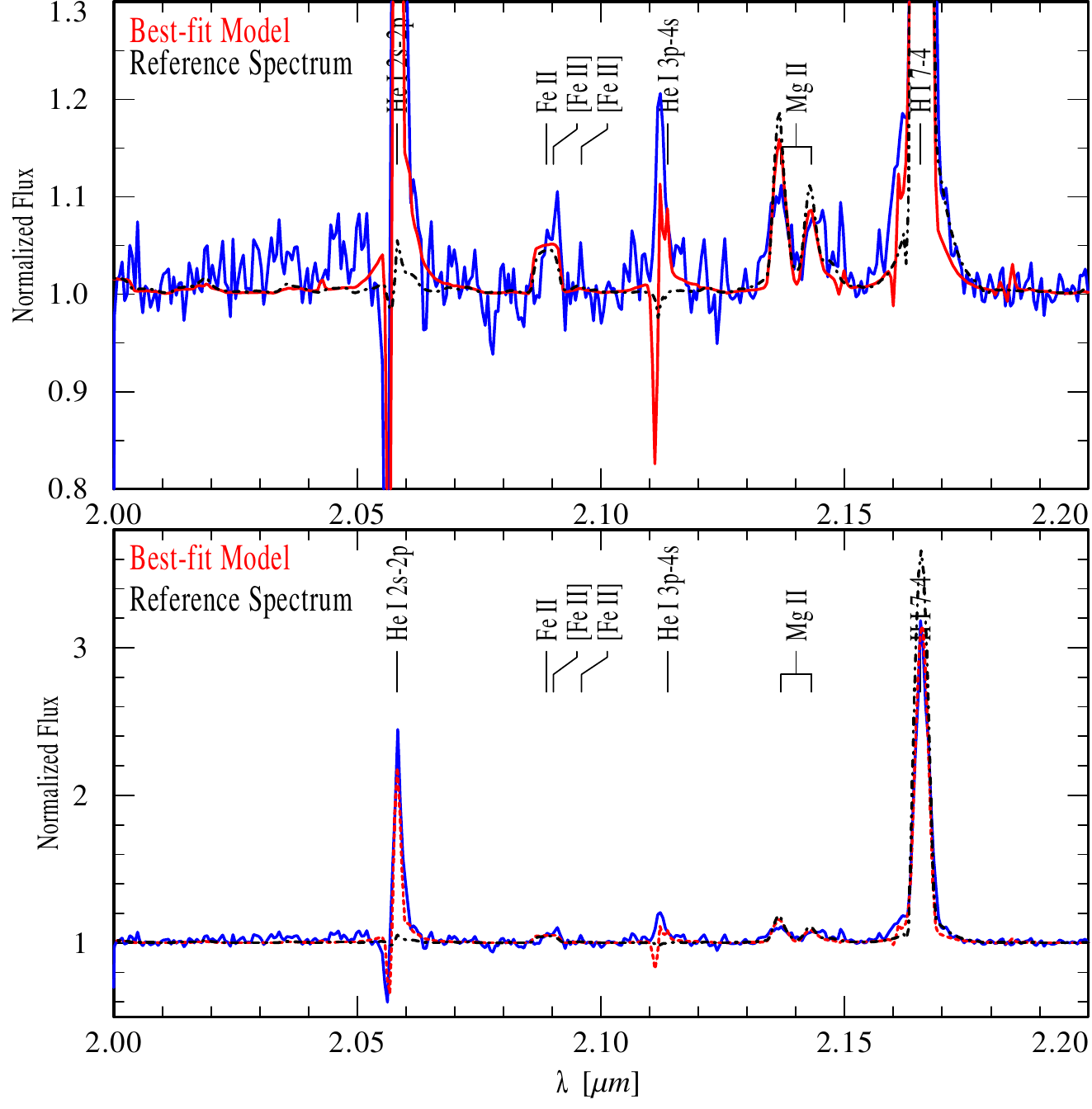}
\caption{Lower panel: GRAVITY spectrum from 2.0
  $\mu$m to 2.2 $\mu$m (blue-solid line) with a vertical scale that
  displays the complete line profiles. Upper panel: magnification of the GRAVITY spectrum, where the fine details of our
  best-fit \texttt{CMFGEN} model (red-dashed line) and of the reference
  spectrum derived with the stellar parameters from \citet{Groh_2012a} (black-dashed line) are
  appreciated. Notice that the He {\sc i} 2s-2p and He {\sc i} 3p-4s are only matched by our hotter red model.}
\label{fig:CMFGEN_model_fit} 
\end{figure}

\begin{figure}[ht]
\centering
\begin{minipage}[t]{1.0\columnwidth}
\includegraphics[width=\columnwidth]{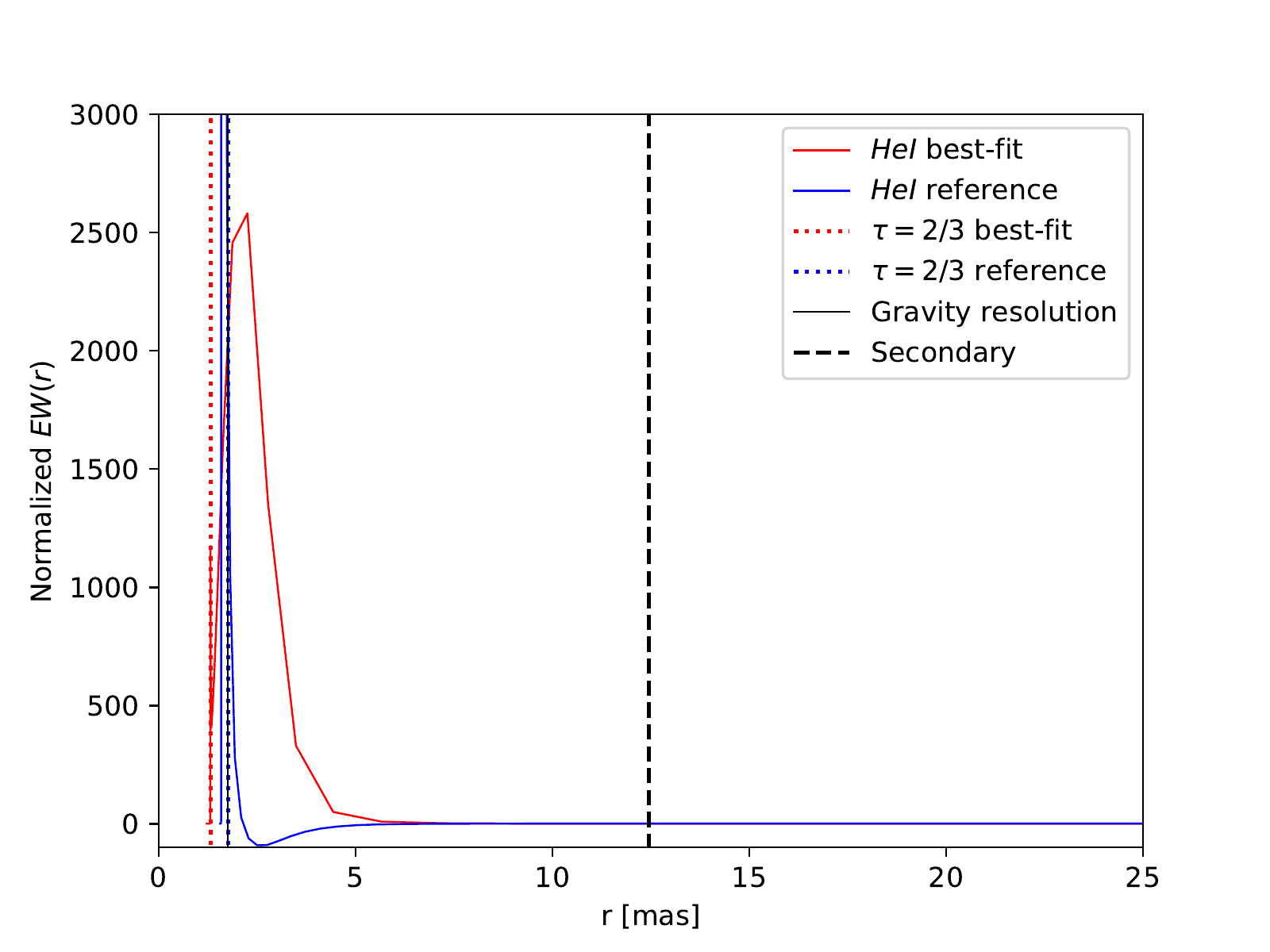}
\end{minipage}
\begin{minipage}[t]{1.0\columnwidth}
\includegraphics[width=\columnwidth]{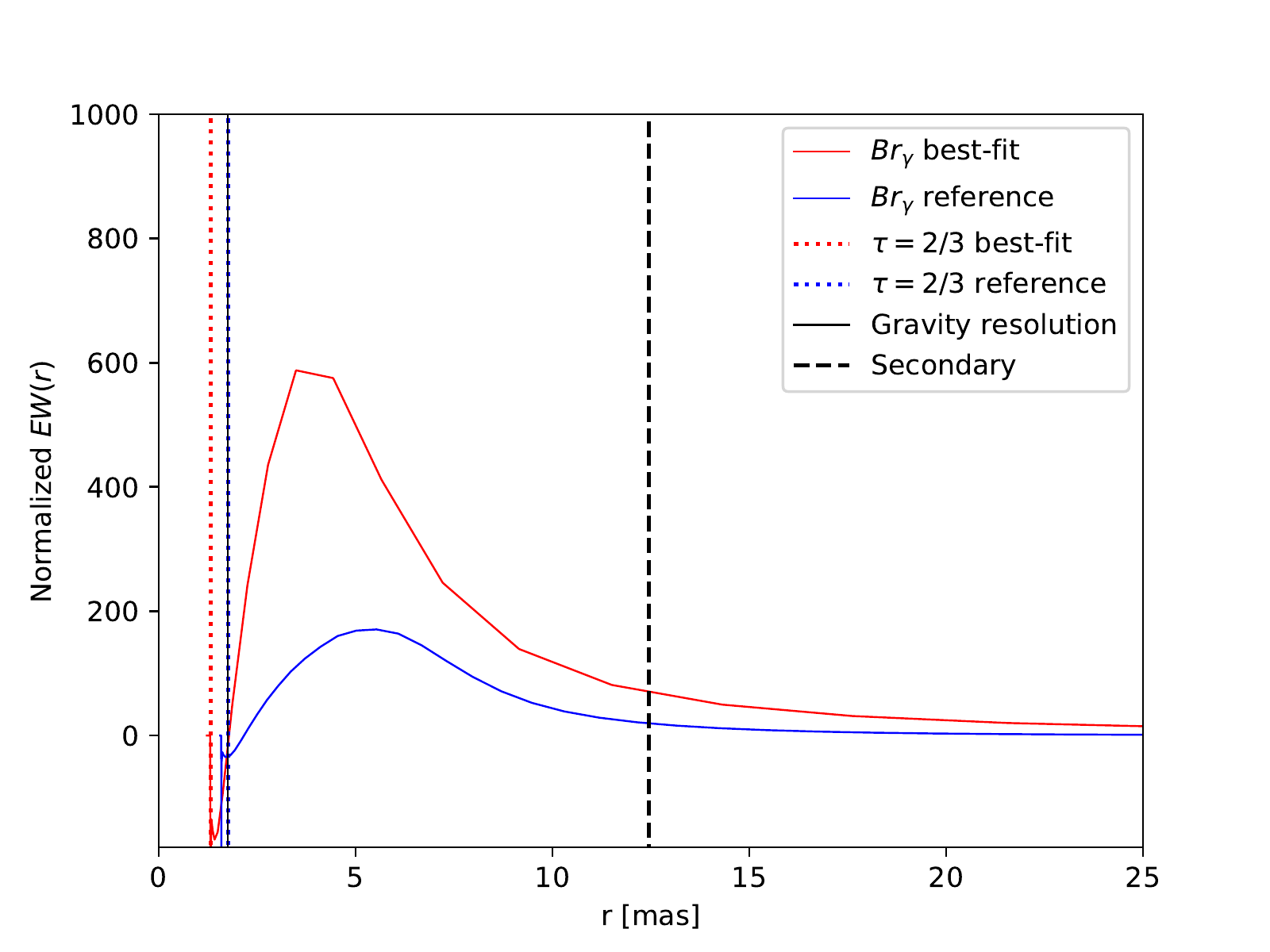}
\end{minipage}

\caption{\texttt{CMFGEN} normalized line equivalent width (EW) as a function of
  radius for our best-fit model and the reference one. Top and bottom panels: line-forming region as a
  function of radius for He {\sc i} 2s-2p and Br$\gamma$, respectively. The red/blue dotted line corresponds the location of the $\eta_A$'s
  photosphere. The black-solid line indicates the spatial
  resolution limit of GRAVITY and the black-dashed line shows the projected
  position of $\eta_B$ at apastron, according with the mean orbital
  solution presented by \citet{Teodoro_2016}. }
\label{fig:CMFGEN_lpf} 
\end{figure}

\subsection{Spectroscopic analysis of $\eta$ Car's primary wind \label{sec:spec_discussion}}

To characterize the properties of $\eta_A$'s wind, we fitted the
integrated spectrum of our GRAVITY observations
(Sect.\,\ref{sec:spectrum}) with model spectra. The synthetic spectra were computed with the 1D spherical non-LTE stellar atmosphere and radiative transfer code \texttt{CMFGEN}
\citep{Hillier_1998}. Our atomic model includes a large number of ionization stages to cover a wide
temperature range in our stellar atmosphere grid without changing the
atomic model. The following ions were taken into account: H\,{\sc i}, He\,{\sc i-ii}, C\,{\sc i-iv}, N\,{\sc i-iv},
O\,{\sc i-iv}, Ne\,{\sc i-iv}, Na\,{\sc i-iv}, Mg\,{\sc i-iv}, Ca\,{\sc i-iv},
Al\,{\sc i-iv}, Si\,{\sc i-iv}, P\,{\sc ii-v}, S\,{\sc i-v}, Ar\,{\sc i-iv}, Fe\,{\sc
  i-v}, and Ni\,{\sc ii-v}; and the metallicity
was set to solar according to \citet{Asplund_2005}. For comparison, we computed a reference spectrum with the stellar parameters derived by
\citet{Groh_2012a} for the primary, which in turn is based on previous estimates by
\citet{Hillier_2001, Hillier_2006}. The terminal
velocity ($\varv_{\infty}$), beta-type velocity law ($\beta$) and
volume-filling factor ($f_{\rm v}$) were taken from \citet{Groh_2012a}
and were not varied for the grid computation. The luminosity was
scaled to match the $K$-band flux of the reference spectrum. The stellar parameters
of our best-fit model and the reference spectrum are listed in
Table\,\ref{table:CMFGEN_parameters}.

While our best-fit model reproduces the spectral lines in the GRAVITY
spectrum between 2.0 $\mu$m and 2.2 $\mu$m, the reference spectrum
fails to reproduce the He\,{\sc i} lines (Figure \ref{fig:CMFGEN_model_fit}). Our best-fit model is about 4\,000
$K$ hotter than the reference spectrum, but the mass-loss rate is
decreased by a factor of two. There is a large debate in the
literature whether $\eta_A$'s mass-loss rate has
decreased by a factor between two to four over the last two decades \citep[see
e.g., ][]{Corcoran_2010, Mehner_2010,
  Mehner_2012, Mehner_2014}. This could explain the change in the
mass-loss estimate from the 2000 HST data used by \citet{Groh_2012a} and our 2016--2017
GRAVITY observations, but not the increased temperature. 

However, this scenario is not fully consistent with
all the observational evidence and theoretical predictions. The hydrodynamic simulations of the wind-wind collision
zone, created by \citet{Madura_2013}, suggest that, instead of an extreme decrease of the $\eta_A$
mass-loss rate, changes in $\eta$ Car's ultra-violet, optical, and
X-ray light curves, as well as of the spectral features, are due to slight changes in the wind-wind collision
cavity opening angle in combination with a moderate change in the
mass-loss rate. 

Moreover, when comparing our \texttt{CMFGEN} model
with our complementary FEROS optical spectrum ($\lambda$4\,200--7\,100 \AA), we noticed that it is not
able to reproduce the metallic [Fe {\sc ii}] and Si {\sc ii} lines,
while the cooler Groh model reproduces them (see Fig.\,\ref{fig:FEROS_spectrum}). Similar
behavior is observed for other previous models that include low primary mass-loss
rates close to $\sim$10$^{-4}\, M_{\odot}$ yr$^{-1}$ \citep[see e.g.,
Fig.\,12 in ][]{Madura_2013}. This speaks against the
strong decrement in the mass-loss rate obtained by our
\texttt{CMFGEN} model, and highlights that $\eta_B$ plays a major role in the formation of the
observed lines in the GRAVITY spectrum. The effects of the secondary
and the wind-wind collision zone on the formation of Br$\gamma$ and
He {\sc i} cannot
be reflected by our 1D non-LTE single star model. Nevertheless,
they can be investigated with aperture-synthesis images.

For a better understanding of the system's geometry and how
$\eta_B$ could change the parameters derived by our \texttt{CMFGEN} model, we show in
Fig.\ref{fig:CMFGEN_lpf} the line-forming regions
of our best-fit model and of the reference one
as a function of radius for He\,{\sc i}
2s-2p (top panel) and Br$\gamma$ (lower panel). The vertical black lines indicate the following
reference points: (a) the red-dotted line shows the location of photosphere
of $\eta_A$, which is not equal to the hydrostatic radius (because
of its optically thick wind) but to the point where the optical depth
$\tau = 2/3$; (b) the black-solid line shows the mean angular resolution limit of
the GRAVITY observations ($\sim$ 2 mas); (c) and the black-dashed line
marks the expected projected
position of
$\eta_B$ at the time of apastron. 

With our \texttt{CMFGEN} model, the wind of $\eta_A$ could form the peak of the Br$\gamma$
emission only at 3 mas
(7 au), which is clearly closer than the extended emission
observed in the reconstructed images. Furthermore,  it can be seen
that, with a single star, the peak of the 
He\,{\sc i} 2s-2p line is formed at an angular scale close the
resolution of our observations (in the reference model, the He {\sc i}
line-formation region is at a scale smaller than the resolution of
GRAVITY). Therefore, the extended emission in the He\,{\sc i} 2s-2p images should
be related to the UV ionization of $\eta_B$ on the
wind-wind collision zone. Any modification to He\,{\sc i}
as a consequence of the wind-wind interaction would require that (i) $\eta_B$'s wind penetrates deeply into the denser regions of the primary
wind and/or (ii) that $\eta_B$ ionizes part of the pre- and
post-shock primary wind near
the apex of the wind-wind collision zone. In this scenario, it is expected
that the modification of the ionization structure caused by the secondary largely changes the intensity of
Helium and Hydrogen lines depending on the orbital phase, particularly at
periastron \citep[see
e.g.,][]{Groh_2010, Richardson_2015, Richardson_2016}.   

We do not detect any significant differences in the line profiles between the 2016 and 2017
GRAVITY spectra. However, when comparing the 2016 GRAVITY data and the
2004 AMBER data reported by \citet{Weigelt_2007} (both of them obtained at a similar orbital phase), we
noticed that the He {\sc i} 2s-2p shows the P-Cygni profile, but the amplitudes of the valley
and peak are different for both data sets. This observation suggests
the presence of a dynamical wind-wind
collision environment with possible changes in the wind parameters
over time \citep[see e.g., Fig.\,3 in ][ where the P-Cygni profile of He
{\sc i} $\lambda$5876 is observed close to several periastron passages
but the profile of the line varies with time even for similar phases]{Richardson_2016}. Therefore, this highlights the importance of monitoring the
lines' morphological changes through the reconstructed interferometric
images. 

\begin{table}
\caption{Stellar parameters of the \texttt{CMFGEN} models.}
\label{table:CMFGEN_parameters}
\centering
\begin{tabular}{c c c} 
\hline\hline
Parameter & Groh et al. (2012) & This work \\
\hline
$T_{\rm eff}$						& $9.4$\,kK			& $13.5$\,kK\\
$\log (L/L_{\odot})^{\mathrm{a}}$				& $6.7$				& $7.1$\\
$Y^{\mathrm{b}}$									& $0.55\%$			& $0.50\%$\\
$\log (\dot{M}/M_{\odot}/{\rm year})$	& $-3.1$			& $-3.4$\\
$\beta^{\mathrm{c}}$								& 1.0				& 1.0\\
$f_{\rm v}^{\mathrm{d}}$							& 0.1				& 0.1\\
$\varv_{\infty}^{\mathrm{e}}$					&$420$\,$\kms$ 		& $420$\,$\kms$\\
\hline
\end{tabular}
\begin{list}{}{}\footnotesize
\item[$^\mathrm{a}$] $\log (L/L_{\odot})$ is scaled to match the $K$-Band flux\,
\itemsep0em
\item[$^\mathrm{b}$] Helium abundance in mass fraction\,
\itemsep0em
\item[$^\mathrm{c}$] Exponent of the wind-velocity law\,
\itemsep0em
\item[$^\mathrm{d}$] Volume filling factor\,
\itemsep0em
\item[$^\mathrm{e}$] Stellar wind's terminal velocity\,
\itemsep0em
\item[*] Note: $\beta$,
  $f_{\rm v}$, and $\varv_{\infty}$ were fixed parameters in our grid
  of models. The used values were taken from \citet{Groh_2012a}
\end{list}
\end{table}

\subsection{The size of $\eta$ Car's continuum emission}

The observed quiescent continuum emission in $\eta$ Car is caused by
extended free-free and bound-free emission \citep{Hillier_2001}, and it traces the dense,
optically thick primary wind. From our
geometrical model to the Fringe Tracker data, we estimated that
$\sim$50\% of the $K$-band total flux corresponds to
$\theta_{\mathrm{FWHM}}$ = 2
mas ($\sim$ 5 au) core (compact
emission), with the rest of the
flux arising from a surrounding halo of at least 
$\theta_{\mathrm{FWHM}}$ = 10 mas ($\sim$ 24 au; see Table\,\ref{tab:best_fit_model}). The size derived for the continuum $\eta_A$ wind is
consistent with the size of the extended primary photosphere, according with
our \texttt{CMFGEN} model and previous spectroscopic and
interferometric estimates \citep{Hillier_2001, Hillier_2006,
  Kervella_2002, vanBoekel_2003, Weigelt_2007, Weigelt_2016}. 

\begin{table*}
\caption{Values of $\eta$ Car's continuum elongation ratio from different interferometric observations}
\label{tab:geom_models}
\centering
\begin{tabular}{c c c c} 
\hline\hline
Reference & Instrument & Elongation ratio & PA \\
\hline
\citet{vanBoekel_2003} & VINCI	& 1.25 $\pm$ 0.05 & 134$^{\circ} \pm$ 7$^{\circ}$\\ 
\citet{Weigelt_2007} & AMBER & 1.18 $\pm$ 0.10 & 120$^{\circ} \pm$ 15$^{\circ}$\\
\citet{Weigelt_2016} & AMBER & 1.07 $\pm$ 0.14 & 159$^{\circ} \pm$ 47$^{\circ}$\\
This work & GRAVITY & 1.06 $\pm$ 0.05 & 130$^{\circ} \pm$ 20$^{\circ}$\\
\hline
\end{tabular}
\end{table*}

Previous interferometric studies suggested an elongated
continuum primary wind along the PA of the Homunculus. From our data modeling, we derived an elongation ratio of $\epsilon$ =  1.06 $\pm$
0.05, a value that is consistent with the most recent interferometric
measurements with AMBER, but it is 4-$\sigma$ smaller than the first
VINCI estimate in 2003 (see Table.\,\ref{tab:geom_models}). It has been
hypothesized that the origin of this elongation could be (i) a
latitude-dependent wind caused by a rapid rotation of the primary
and/or (ii) a consequence of the wind-wind collision cavity. For example, \citet{Mehner_2014} suggest that the angular
momentum transfer between $\eta_A$ and $\eta_B$ at periastron
could be affected by tidal acceleration, resulting in a change of the
rotation period of $\eta_A$, which in turn may affect the shape of the continuum emission.

Additionally, \citet{Groh_2010a} used radiative transfer models
applied to the VINCI and AMBER data to explore the changes in the
continuum associated with modifications in the rotational velocity
and with a ``bore-hole'' effect due to the presence of the wind-wind
collision zone. These authors found that both prolate or oblate rotational
models reproduce the interferometric signatures. Nevertheless,
those models required large inclination angles in which $\eta_A$'s
rotational axis is not aligned with that of the Homunculus. Inclinations where the rotational axis is aligned with the
Homunculus would require a decrement in the rotational velocity (and of the
size) of $\eta_A$ with time. However, the cavity could also mimic the effects of the observed
elongation. Therefore, new radiative transfer models of interferometric data
(including GRAVITY and other wavelengths like PIONIER in the $H-$band)
are necessary to test these scenarios. Such models are beyond
the scope of the present work and are left for future analysis. 

\begin{figure}[htp]
\centering
\includegraphics[width=\columnwidth]{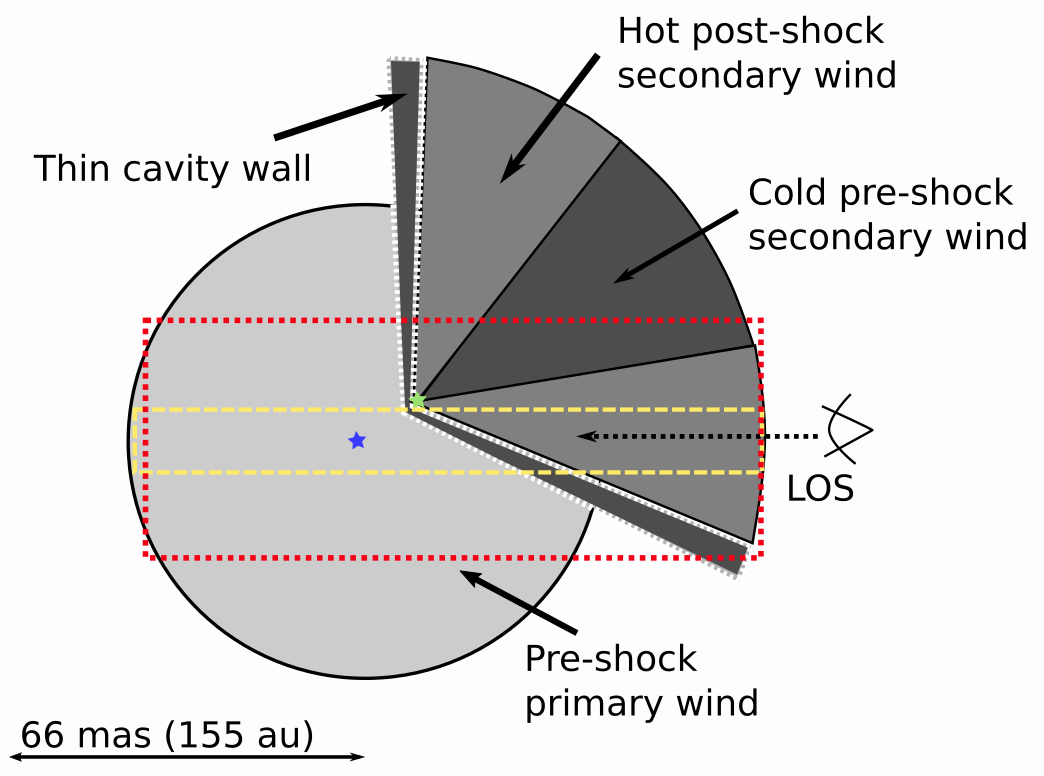}
\caption{Schematic of $\eta$ Car's wind-wind collision
  scenario reported by \citet{Madura_2013}. The wind-wind collision cavity carved by the
secondary (green star) in the primary wind is depicted, with the
different elements of the primary (blue star) and secondary winds labeled. Notice how
the LOS lies preferentially toward the lower cavity wall. The angular
scales (centered around $\eta_A$) traced with the shortest baselines of the
GRAVITY ($\sim$40 m) and AMBER ($\sim$10 m) data are shown with a yellow-dashed and red-dotted rectangles, respectively. }
\label{fig:EtaCar_BrG_diagram} 
\end{figure}

\subsection{The Br$\gamma$ interferometric images in
  the context of the $\eta$ Car wind-wind collision cavity \label{sec:physical_model_BrG}}

\begin{figure*}[htp]
\centering
\includegraphics[width=\textwidth]{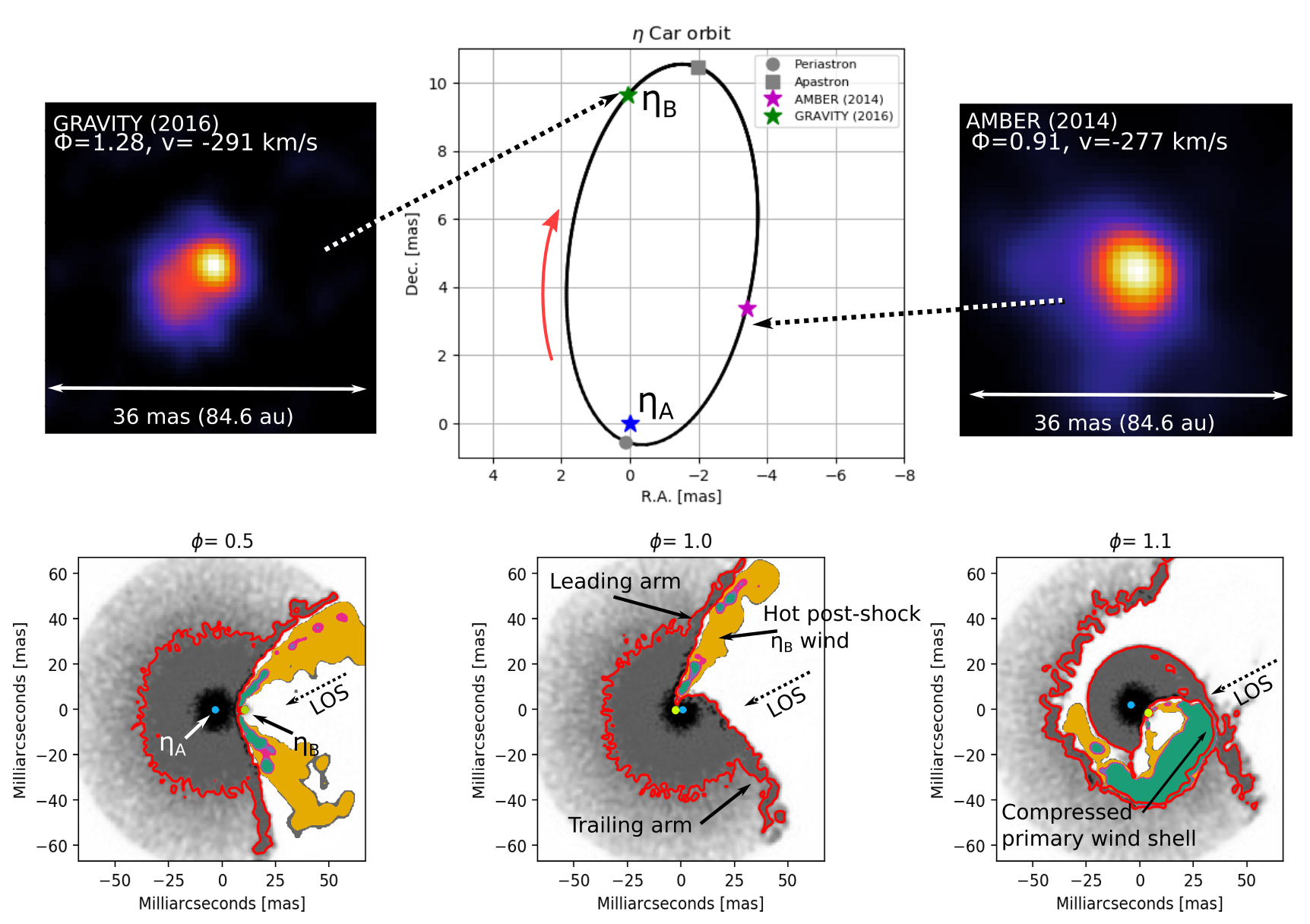}
\caption{Upper-middle panel: projected orbit of $\eta_B$ around $\eta_A$, with the
position of the periastron, apastron, and two imaging epochs
labeled on it. Upper-right panel: AMBER map at $-$277 $\kms$, where the
  fan-shaped SE morphology is observed at $\phi \sim$ 0.91. Upper-left
  panel: GRAVITY image at $-$291 $\kms$, the SE arc-like
  feature is observed at $\phi \sim$ 1.28. The FOV of the
images is of 36 mas, and they are oriented with the east pointing toward the left
and the north toward the top of frames. Lowermost panels: three orbital phases, $\phi$=0.5 (lower-left), $\phi$=1.0
(lower-middle) and $\phi$=1.1 (lower-right), of the wind-wind collision model created by
\citet{Madura_2013}. The panels show the structure of the wind-wind
cavity in the orbital plane. The red lines indicate the contour where
the density is 10$^{-16}$ g/cm$^3$ and they highlight the
changes in the leading and trailing arms of the wind-wind cavity. The positions of the primary and secondary for
each phase are marked on the panels with blue and green dots,
respectively. The green, magenta, and yellow areas trace the hot ($T\,\sim$ 10$^{6}$ - 10$^{8}$ K)
post-shock secondary wind bordering the cavity shells. The vector of the observer's line-of-sight is labeled on
each panel, and it corresponds to $\omega =$243$^{\circ}$ and $\Omega
=$47$^{\circ}$. The principal components of the
wind-wind cavity are also indicated
in the panels. The
lowermost panels were taken from \citet{Madura_2013} and adapted to
the discussion presented in
Sect.\,\ref{sec:physical_model_BrG}. }
\label{fig:ims_comparison} 
\end{figure*}

Our 2016 GRAVITY Br$\gamma$ aperture-synthesis images reveal the
morphology of $\eta$ Car's wind-wind collision cavity. Here, we discuss the principal wind components and their interpretation with
previous theoretical models and simulations. The
Br$\gamma$ maps reveal the following general structure as function of the radial
velocity:

\textit{ (1) Continuum wind region.} For radial velocities smaller
than $-$510 $\kms$ and larger than $+$368 $\kms$, the maps show a compact emission where the optically thick continuum
  wind region is dominant. For velocities between $-$510 $\kms$ and $+$368 $\kms$, additional wind components are observed in
  the images, however, the continuum emission is always present.

\textit{(2) Wind-wind collision region.} 3D smoothed particle
hydrodynamic models \citep{Okazaki_2008, Gull_2009, Madura_2012b, Madura_2013,
  Teodoro_2013, Teodoro_2016} of $\eta$ Car's wind-wind
interaction suggest a density distribution that extends far beyond the size of the primary
wind. This wind-wind collision zone is identified with a cavity
created by the fast wind of $\eta_B$ that penetrates deep into the
dense but slow wind of $\eta_A$ as the secondary changes its
orbital phase. Following the orbital solution of \citet{Teodoro_2016} presented
in Fig.\,\ref{fig:EtaCar_orientation}, $\eta_B$ was close to apastron
(in front of $\eta_A$) at the time of our GRAVITY
observations. Figure \ref{fig:EtaCar_BrG_diagram} displays a schematic
view of this wind-wind
scenario according to \citet{Madura_2013}. The diagram represents the
position of the system at apastron in a plane
defined by the LOS and the sky plane \citep[i.e., called the xz plane
in][]{Madura_2013}. The cavity opens with a LOS oriented
preferably toward the southern wall, along its walls the hot ($T\,\sim$ 10$^6$ K) post-shock
$\eta_B$ wind is moving, and, in between, the cold ($T\,\sim$ 10$^4$ K)
pre-shock $\eta_B$ wind is observed. 

In the Br$\gamma$ GRAVITY images at radial velocities > $-$364 $\kms$, we observe an
  extended asymmetric emission increasing in size and brightness as we
  approach the systemic velocity, and then it decreases toward
  positive ones. At velocities between $-$364 $\kms$ and $-$71
  $\kms$, the wind-wind collision cavity presents an elongated and asymmetric
  cone-like morphology with a bright arc-like feature in the
  southeast direction. The extended and bright southeast asymmetric emission was first
observed in the AMBER images reported by
\citet{Weigelt_2016}. In particular, for velocities between $-$140 $\kms$
and $-$380 $\kms$, those authors
identified a SE fan-shaped morphology. However,
there are some changes between the AMBER
images and the maps recovered with
GRAVITY. We infer that some of those changes are associated with the
dynamics of the wind-wind collision zone (i.e., different orbital phases), however, some others are caused by the different spatial frequencies
sampled  with the two observations. 

The majority of the longest baselines of the AMBER data are of 80 m, which limits the angular resolution up to
$\lambda$ /2B$_{\mathrm{max}}$ = 3 mas, in
  contrast our GRAVITY data provide a mean maximum angular resolution
  of 2.26 mas. However, the AMBER
data include a better sampling of low spatial frequencies with
baselines as small as 10 m. In this respect, our
GRAVITY data present a clear limitation since most of our short
baselines are above 40 m. Those constraints allowed us to
recover images of the small-scale structure of
$\eta$ Car, at the cost of restricting the imaging capabilities to map the most extended
emission observed in the AMBER images. 

\citet{Madura_2013} suggested that the time-dependent changes in the
spectral lines are a consequence of changes in the line-of-sight
morphology of the wind-wind collision cavity.  Their
hydrodynamical simulations predict that, at apastron, the cavity maintains an axisymmetric conical shape with the
leading and trailing arms clearly visible in the orbital plane \footnote{For a complete description of
  the changes in the wind-wind collision cavity and their 3D
  orientation at scales relevant to the interferometric images, we
  refer the reader to Fig.\,1, 2, B1, B2, B3 and B4 in
\citet{Madura_2013}.}. The half-opening angle of the cavity depends
on the ratio of the wind momenta, which can be expressed in terms of
mass-loss rate and wind velocity ($\beta$ = $\dot{M_{\eta_A}}\nu_{\eta_A} /
\dot{M_{\eta_B}}\nu_{\eta_B}$), of the two interacting
stars. Therefore, any modification to the mass-loss rates and/or wind
velocities would change the opening angle of the cavity. For example,
$\dot{M}_{\eta_A}$ = 8.5$\times$10$^{-4}\,M_{\odot}$ yr$^{-1}$ would produce a half-opening
angle close to 55$^{\circ}$. On the other hand, $\dot{M}_{\eta_A}$ =
1.6$\times$10$^{-4}$ $M_{\odot}$ yr$^{-1}$ would create a half-opening
angle close to 80$^{\circ}$. 

\begin{figure*}[htp]
\centering
\includegraphics[width=\textwidth]{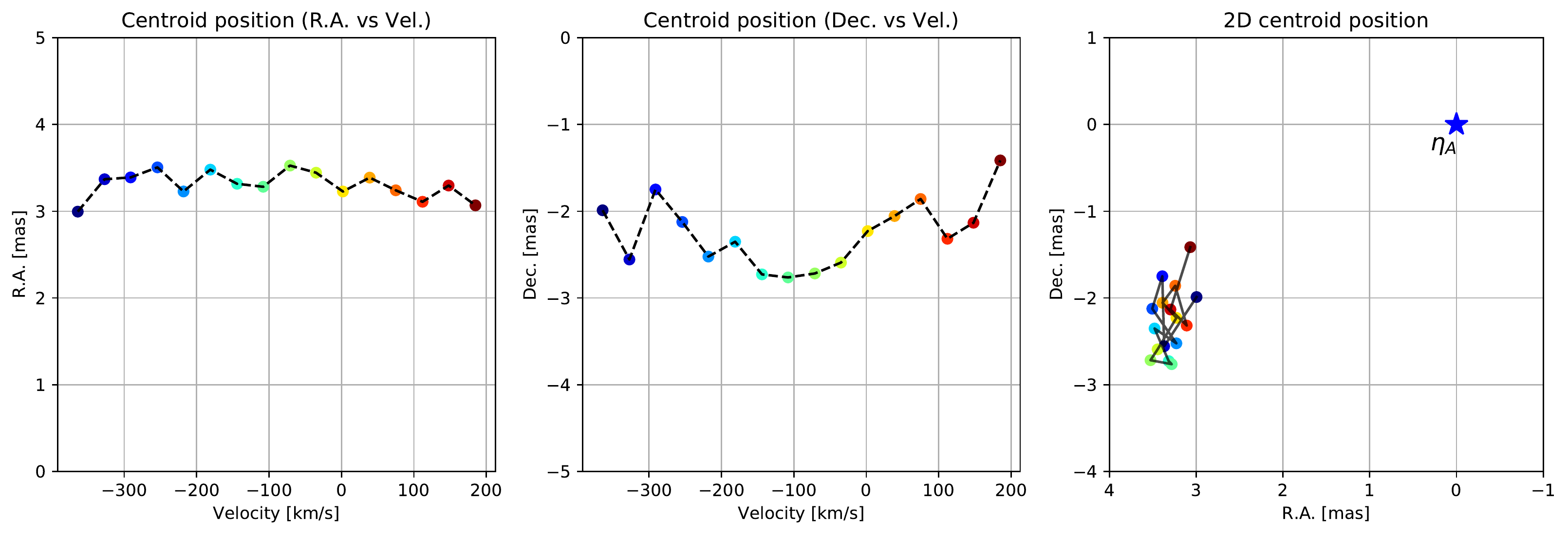}
\caption{The image displays the centroid position of the 2016 arc-like
  feature vs velocity. Left panel: centroid
  changes in R.A vs velocity. Middle panel: centroid changes in
  Dec. vs velocity. Right panel: 2D centroid position relative to the $\eta_A$'s continuum marked at origin
  with a blue star.}
\label{fig:3d_projection} 
\end{figure*}

As $\eta_B$ moves from apastron (lower-left panel in
Fig.\,\ref{fig:ims_comparison}) toward the periastron, the wind-wind cavity is
distorted, creating a spiral (lower-middle panel in
Fig.\,\ref{fig:ims_comparison}). Interesting is the phenomenology after the
periastron passage, where the leading arm of the cavity collides with
the ``old'' trailing arm, forming a dense, cold, and compressed ``shell'' of $\eta_A$'s wind (lower-right panel in
Fig.\,\ref{fig:ims_comparison}). Since the secondary wind is
  providing pressure from the direction of the binary, but because of
  the very high velocity, the shell is expanding outward through the
  low-density region
on the far (apastron) side, the expansion velocity could reach
values larger than the terminal velocity of $\eta_A$'s
wind ($\nu_{\infty}$=420 $\kms$). The stability and size of the
compressed $\eta_A$ shells are
also dependent on the adopted $\dot{M}_{\eta_A}$. Bordering the
compressed shell of primary wind, there is the hot
post-shock $\eta_B$ wind. Due to the increment of orbital speed
of $\eta_B$ as it approaches the periastron, the post-shock $\eta_B$
wind is heated to higher temperatures ( $T\,\sim$ 10$^{6}$ K) than the gas in the trailing
arm ($T\,\sim$ 10$^{4}$ K), producing an asymmetric hot shock. After
the periastron, photo-ablation of the post shock $\eta_A$ wind could stop
the $\eta_B$ wind from heating the primary wind, enhancing the
asymmetric temperature in the inter-cavity between the secondary and (bordering) the
compressed shell of primary wind. 

In this framework, the 2016 GRAVITY images reveal the structure of
the cavity at an orbital phase $\phi \sim$ 1.3\footnote{Here, we adopted the 2009 periastron passage,
  T$_0$ =2454851.7 JD, as
  reference for the orbital phases labeled on the reconstructed images
  and diagrams of the wind-wind cavity structure.} (upper-left panel
in Figure\, \ref{fig:ims_comparison}), where the arc-like feature could be interpreted as the inner-most hot post-shock gas
flowing along the cavity walls that border the compressed $\eta_A$ wind shell
in our line-of-sight. The fact that we observe material preferentially
along the SE cavity walls, suggests that the half-opening angle is
quite similar to the inclination angle of the system.

\citet{Madura_2013} shows detailed model snap-shots that illustrate
the dependence of the opening angle on the mass-loss rate. However, a
straight-forward estimate following the two-wind interaction solution
of \citet{Canto_1996}, keeping constant the secondary mass-loss rate
and wind velocities, suggests that a primary mass-loss rate between 5.8
$\times$ 10$^{-4}$ and 1.2 $\times$ 10$^{-3}$ $M_{\odot}$ yr$^{-1}$ is necessary to produce a half-opening angle
in the cavity of 40$^{\circ}$--60$^{\circ}$. Smaller mass-loss rates
would produce bigger angles, which would prevent the wall of the
wind-wind collision cavity from intercepting the LOS. The orientation
of the extended emission is also consistent with a
prograde motion of the secondary (i.e., $i$ > 90$^{\circ}$, $\omega$ >
180$^{\circ}$), with a leading arm projected motion from east to west. 

Our analysis of the Br$\gamma$ arc-like feature's peak in the 2016 images suggests
that not all the material is moving at the same speed,  Figure \ref{fig:3d_projection}
displays the centroid position (flux centroid of the 90\%
emission peak) of the arc-like feature versus radial velocity. As
it is observed, the centroid distribution varies between 3.0 and 4.0 mas (7 to 9 au) and between $-$1.5 and $-$3.0 mas
(3.5 to 7 au) in R.A. and Dec.,
respectively. This indicates that the observed
material has a clumpy structure that is not expanding at
the same speed. It is also
consistent with the hydrodynamic simulations which suggest that the
stability of the shells depends on the shock thickness and that fragmentation begins later the higher the value of the primary's
mass-loss rate. 

Furthermore, it also provides a plausible explanation for the
structures observed in the 2014 AMBER images \citep{Weigelt_2016}, which correspond to a
later orbital phase previous to the periastron passage (upper-right panel in Fig.\,\ref{fig:ims_comparison}), where the shells are disrupted and the hot
$\eta_B$ wind has already mixed with some portions of the compressed
$\eta_A$ wind. In this scenario the bar, observed in the AMBER data between $-$414 and
$-$352 $\kms$, could be part of the highest velocity component of the hot
gas around the distorted shell, while the two antennas of the fan-like
structure (from $-$339 to $-$227 $\kms$) could trace portions of the shell
that are not yet fragmented. The SE extended emission at lower
negative velocities could be interpreted as part of the material bordering the
disrupted shell. The observed extended SW emission at positive
velocities could be identified as the material flowing along the
leading arm's wall which is moving away from the observer at the
orbital phase of the AMBER images. 

This scenario agrees with observations of the fossil shells on larger spatial scales. \citet{Teodoro_2013}  observed
three progressive shells using HST [Fe {\sc ii}] and [Ni {\sc ii}]
spectroimages. Since the emission of the forbidden lines is
  optically thin, the observed shells in the [Fe {\sc ii}] and [Ni
  {\sc ii}] images support a future radio mapping with ALMA in H$_{\alpha}$ lines at an angular resolution better than
  0.1'' to properly constrain their emmitting regions. The positions of the arcs are consistent with the SE extended
emission observed in the interferometric images, but the HST arcs
extend up to 0.5''. Those authors derived a time difference between
the arcs of the order of
the orbital period ($\sim$5.54 years), suggesting that each one was
created during a periastron
passage. Assuming a constant shell velocity, \citet{Teodoro_2013}
derived a traveling speed of the fossil shells of $+$475 $\kms$, with the closest
arc to $\eta_A$ located, in projection, at $\sim$0.1'' (235
au). Notice that the innermost arc presented in Fig.\,2 of
\citet{Teodoro_2013} resembles the fan-like structure observed in the
AMBER images.

It is important to highlight that, while the HST images
are tracing the compressed $\eta_A$ wind in the fossil shells, our Br$\gamma$
images are tracing the hot gas, bordering them, along the cavity
walls. A monitoring of the arc-like features with
GRAVITY (using short baselines) will be fundamental to
characterize the primary's mass-los rate and the effect of $\eta_B$
over the primary wind shells over future orbital phases.

\begin{figure}[htp]
\centering
\includegraphics[width=6 cm]{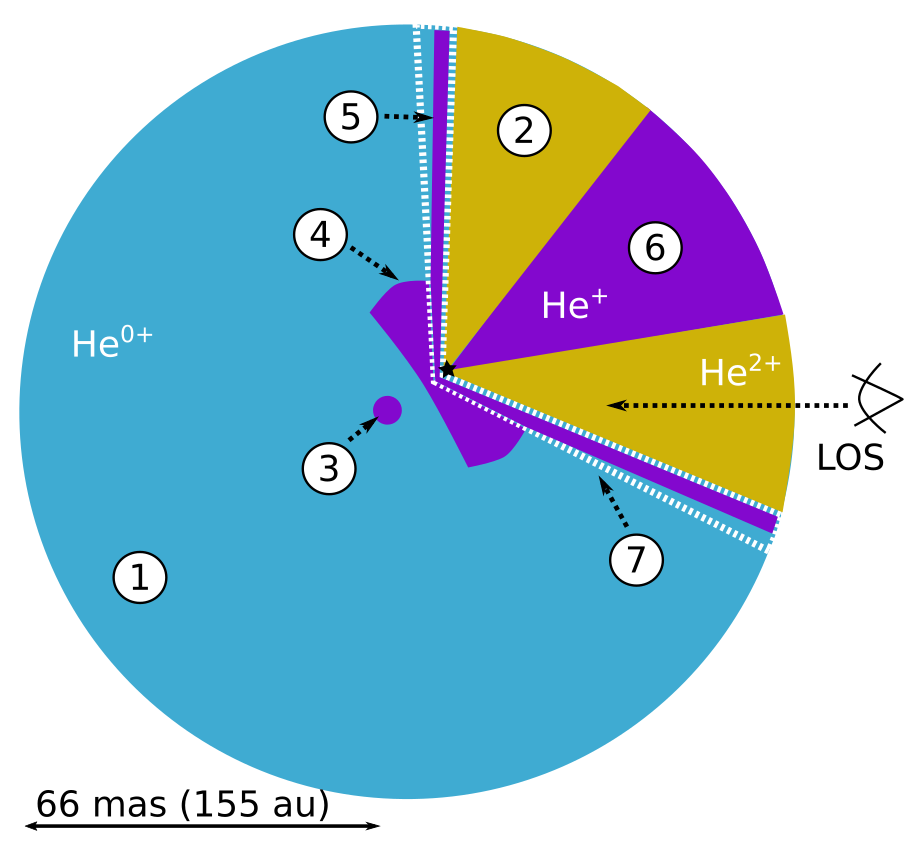}
\caption{Different He ionized
  regions: He$^{0+}$ (blue), He$^{+}$ (purple), He$^{2+}$ (yellow). The
sketch is oriented in a plane parallel to the
line-of-sight and the sky plane. Different regions in the wind-wind collision scenario
are labeled. Region 1 represents the He$^{0+}$ zone in the primary
wind. Region 2 shows the hot post-shock secondary wind composed by
He$^{2+}$. Regions 3, 4, 5, and 6 correspond to the zones composed by
He$^{+}$ in the primary and secondary winds. Region 7 displays the borders of the cavity walls. }
\label{fig:He_diagram} 
\end{figure}
 
\subsection{Ionization effect of $\eta_B$'s wind observed through
  the He {\sc i} 2s-2p line} 

Helium lines are one of the most intriguing spectral features in $\eta$ Car's
spectrum. For example, the discovery of the binary was in part based on spectroscopic
variations of the H-band He {\sc i} lines \citep{Damineli_1996, Damineli_1997}. Since then, several authors have studied the variations in
the helium profiles as a consequence of a binary signature \citep{Groh_2004, Nielsen_2007, Humphreys_2008, Damineli_2008a, Mehner_2010,
Mehner_2012}. The different \texttt{CMFGEN}
models support that the helium lines could be originated in the
densest wind of $\eta_A$. Depending on the
$\dot{M}$ used, the line-emitting region could be extended from a few au to
several hundred au \citep[][see also Sec.,\ref{sec:spec_discussion}]{Hillier_2001,
  Hillier_2006, Groh_2012a}. 

However, these models do not fully reproduce the observed
evolution of the He {\sc i} lines without considering the role of
$\eta_B$ \citep[see e.g., ][]{Nielsen_2007, Damineli_2008a}. It is still not clear where and how the Helium lines are formed in
the wind-wind collision scenario. \citet{Humphreys_2008}, based on HST
observations, suggested that the He {\sc i} lines originate at spatial
scales smaller than 100--200 au ($\sim$40--80 mas), but still the structure of the line-formation
region was not indicated. The recent 3D hydrodynamic simulations of \citet{Clementel_2015a,
  Clementel_2015b} presented maps of the different ionization regions
(He$^{0+}$, He$^{+}$, He$^{2+}$)
at the innermost 155 au of $\eta$ Car's core. These simulations take
into account the ionization structure of $\eta_A$'s wind and the
effect of $\eta_B$'s wind on the collision
region for the three different mass-loss rates studied in \citet{Madura_2013}. These models show that
there are several regions composed by He$^{+}$ that might
be responsible for the changes observed in the He {\sc i} profiles. 

The GRAVITY observations presented in this study allowed us to obtain the first
milliarcsecond resolution images of the inner 20 mas (50
au) He {\sc i} 2s-2p line-emitting region. In contrast to Br$\gamma$,
which is observed purely in emission, He {\sc i} 2s-2p shows a P-Cygni profile with the absorption side
blue-shifted. This profile is similar to the helium lines observed at other
wavelengths \citep[see e.g.,
][]{Nielsen_2007}. In our images, at velocities between $-$403 and $-$326 $\kms$, which correspond to the
valley of the line, the extended emission is
brighter than at other wavelengths. As we move from the valley to the
peak of the line, the brightness of the extended emission
decreases, slightly increasing again at velocities around the line's peak
(between $-$95 and $-$18 $\kms$). At the red-shifted
velocities, the continuum is dominant and the extended component is marginally observed at low-velocities. At the valley
of the line, the 2016 images show an elongated emission
that is oriented at a PA $\sim$ 130$^{\circ}$ and it extends on both
sides of the continuum, with the southeast side slightly more
extended. 

To compare the morphological changes observed in the images with the
model developed by \citet{Clementel_2015b, Clementel_2015a}, we use an
adapted version of the helium distribution presented in
Fig. 8 by \citet{Clementel_2015a}. The schematic diagram, in
Fig.\,\ref{fig:He_diagram} displays the distribution of helium, in the
inner 155 au of $\eta$ Car. Region 1 corresponds to the zone in the primary wind that is composed mostly by He$^{0+}$. Region 2 shows the hot
post-shock $\eta_B$'s wind composed mostly by fully ionized
He$^{2+}$. Regions 3, 4, 5, and 6 correspond to the zones composed by
He$^{+}$ and region 7 displays the borders of the cavity walls where
compressed post-shock $\eta_A$'s wind is found.

From the distribution of He$^{+}$, four main regions are observed at
different morphological scales. The first of them (region
3) corresponds to the pre-shock primary wind that is partially ionized
only by $\eta_A$. The second zone (region 4) is created by the
He$^{0+}$-ionizing photons from $\eta_B$'s wind that penetrate the
pre-shock primary wind. This region varies in size, being more compact
along the azimuthal plane than in the orbital
one. \citet{Clementel_2015a} argued as possible cause for these
differences the less turbulent nature of the wind-wind interaction
region in the azimuthal plane. Region 5 shows the He$^{+}$, of the
post-shock wind of $\eta_A$, trapped in the walls of the
cavity. Finally, region 6 displays the section of the pre-shock
$\eta_B$ wind with He$^{+}$. It is important to highlight that regions
3, 4, and 5 are expected to be around three orders of magnitude denser than region 6.

The fact that we observe the absorption side of the He {\sc i} 2s-2p
blue-shifted implies that it is formed by material in front of $\eta_A$. It
means that regions 4, 5, and 6 are in the LOS. Therefore, the He {\sc i} 2s-2p structure observed in the GRAVITY
images is formed by the different contribution of those regions. By
analyzing the images, we hypothesized that the elongated structure, at
velocity channels near the valley of the line, appears to
be formed by the emission coming mainly from regions 4 and 5. This idea is
consistent with the flow of the He$^{+}$ pre-shock primary wind, coming
toward us at a velocity of $\sim$ $-$400 $\kms$, near the apex of the
wind-wind collision zone.

This effect could explain why we observe the most prominent emission
at large velocity channels between $-$403 and $-$326 $\kms$. Given the
PA of the observer's LOS and the orbital plane, we expect to observe
emission coming from region 4 at the top and bottom of the
continuum. This scenario is also consistent with the emission observed
in the GRAVITY images. Since the system near apastron is
preferentially oriented toward the material in region 4 that borders
the lower wall of the cavity, we also expect to observe an asymmetric
emission as the one detected in the images.

As we move toward the systemic velocity of the line, we notice that the extended emission decreases in brightness and size, but it increases again at velocity channels near the emission peak.  We suspect that
this effect is caused by the material in region 5. The post-shock
primary wind in the thin walls of the cavity experiences a turbulent
environment, particularly at the apex, that causes some of the
material to have zero or positive radial velocities.  Notice,
however, that for velocities larger than $+$98 $\kms$, the main source of the
emission is within the size of the primary beam. This suggests that
the origin of large red-shifted velocities is the
pre-shock ionized wind from $\eta_A$ (region 3). Hence, similar to
Br$\gamma$, the observed spatial distribution of He {\sc i} is in
agreement with \citet{Clementel_2015b} models with $\dot{M}_{\eta_A
}$ close to 10$^{-3} M_{\odot}$ yr$^{-1}$. However, future, new hydrodynamic models of the He {\sc i} 2s-2p line in
addition to a GRAVITY monitoring of the line's morphology are
necessary to provide
solid constraints on the mass-loss rate of the system. 

\section{Conclusions \label{sec:conclusions}}

\begin{itemize}   
\item  We present GRAVITY interferometric data of $\eta$
Car's core. Our observations trace the inner 20 mas (50 au)
of the source at a resolution of 2.26 mas. The spectro-interferometric capabilities of GRAVITY allowed us to
chromatically image Br$\gamma$ and He {\sc i} 2s-2p emission regions with a
spectral resolution of R = 4000, while the analysis of the Fringe
Tracker data allowed us to measure the size of the continuum emission,
and the integrated spectrum to characterize the parameters of the primary star.

\item From our geometrical model of the Fringe Tracker squared visibilities, we
  constrained the size of the continuum emission. We derived a mean FWHM
  angular scale of 2 mas ($\sim$ 5 au) and an elongation ratio
  $\epsilon$ = 1.06 $\pm$ 0.5 for the compact continuum emission of $\eta_A$,
  together with an extended over-resolved emission with an angular
  size of at least 10 mas. These estimates are in agreement
  with previous interferometric measurements. Some of the plausible
  hypotheses to explain the observed elongation of the continuum
  compact emission are the fast rotation of the primary and/or the
  effect of the wind-wind collision cavity. A future monitoring of the continuum size and elongation with GRAVITY and
  other interferometric facilities at different wavelengths, in combination with radiative
  transfer models, would serve to test these scenarios.

\item To characterize the properties of $\eta_A$'s wind, we
  applied a \texttt{CMFGEN} 1D non-LTE model to the GRAVITY spectrum. Our
  model reproduces the He {\sc i} 2s-2p and He {\sc i} 3p-4s lines, which could not be formed with the
  parameters used in previous models in the literature. However, the
  line-emitting regions of the best-fit model are quite small compared
  to the structure observed in the reconstructed images. These
  results imply that single-star models are not enough to reproduce
  all the observational data. Therefore, the role of $\eta_B$ should
  be taken into account when modeling
  the $\eta$ Car spectrum.  Furthermore, when
  comparing our model with the $\eta$ Car spectrum in the visible, we
  could not reproduce the observed metallic features. We
    suspect that this is caused because many of the metallic lines are
    forbidden lines with critical electron densities of 10$^{7-8}$
    cm$^{-3}$.  With the FEROS aperture, this emission is originating in the fossil wind shells and hence not reproduced in the stellar model. 

\item Our aperture-synthesis images allowed us to
  observe the inner wind-wind collision
  structure of $\eta$ Car. Previous AMBER interferometric images of
  Br$\gamma$ revealed the wind-wind collision cavity produced by the
  shock of the fast $\eta_B$'s wind with the slow and dense $\eta_A$'s
  wind. Our new 2016 GRAVITY Br$\gamma$ images show the structure of
  such cavity at a different orbital phase with a 2.26 mas resolution. The
  observed morphologies in the images are (qualitatively) in agreement with
  the theoretical
  hydrodynamical models of \citet{Madura_2013}. Particularly
  interesting is the bright SE arc-like feature, which could be
  interpreted as the hot post-shock gas flowing along the cavity
  wall (oriented toward the observer) that border the innermost shell of compressed primary wind, which is
  formed by
  the shock of the cavity's trailing arm with the leading arm after
  the most recent periastron passage. 

\item Due to the sparseness of the 2017 GRAVITY data the quality of
  the reconstructed images from this epoch is clearly
  affected. Therefore, those images could not be used for a direct
  comparison with the 2016 ones. Nevertheless, our analysis of the
  interferometric observables, in coincidental baselines, reveals
  changes in the cavity structure. The ulterior characterization of
  those changes is subject to future imaging epochs with a less
  limited $u-v$ coverage than the 2017 data.

 \item We presented, the first images of the He {\sc i}
   2s-2p line. They were qualitatively interpreted using the model of \citet{Clementel_2015b,
     Clementel_2015a}. We hypothesized that the observed emission is
   coming mainly from the He$^{+}$ at the cavity walls and from a
   portion of the pre-shock primary wind ionized by the secondary. From the size
   of the observed emission and the theoretical models, we suspect
   that the mass-loss rate of the primary is close to 10$^{-3}\,M_{\odot}$ yr$^{-1}$. The emission observed require an active role
   of $\eta_B$ to ionize the material near the apex of the wind-wind
   collision cavity. 

\item Spectro-interferometric imaging cubes offer us unique information to constrain
  the wind parameters of $\eta$ Car, not accessible by other
  techniques. In this study, we have shown the imaging capabilities of
  GRAVITY to carry out this task. A future monitoring of
  $\eta$ Car over the orbital period (particularly at the periastron
  passage), in combination with dedicated hydrodynamical models of the
  imaged $K$-band lines, will provide a unique opportunity to
  constrain the stellar and wind parameters of the target, and,
  ultimately, to predict its evolution and fate. 

\end{itemize}

\begin{acknowledgements}
We thank the anonymous referee for his/her comments to improve this
work. J.S.B acknowledges the support from the Alexander von Humboldt
Foundation Fellowship program (Grant number ESP 1188300 HFST-P) and
to the ESO Fellowship program. The
work of A.M. is supported by the Deutsche Forschungsgemeinschaft
priority program 1992. A.A., N.A. and P.J.V.G. acknowledge funding
from Funda\c{c}\~{a}o para a Ci\^{e}ncia e a Tecnologia (FCT)  (SFRH/BD/52066/2012; PTDC/CTE-AST/116561/2010; UID/FIS/00099/2013) and the European Commission (Grant Agreement 312430, OPTICON).
\end{acknowledgements}

\bibliography{/Users/sanchezj/Documents/Papers/Paper_lib}

\begin{appendix}

\section{GRAVITY observations' lists \label{sec:appA}}

\begin{table*}
\caption{$\eta$ Car's 2016 GRAVITY observations}
\label{tab:2016_obs}
\centering
\begin{tabular}{c l l l c c c c} 
\hline\hline
Date & Source & MJD & Type & DIT & NDIT & Airmass & Seeing \\
\hline
24-02-2016 & $\eta$ Car & 57443.0859 & SCI & 10 & 30 & 1.479 & 0.83 \\
& $\eta$ Car & 57443.0951 & SCI & 10 & 30 & 1.440 & 0.89 \\
& HD\,89\,682 & 57443.1119 & CAL & 10 & 30 & 1.267 & 0.82 \\
& $\eta$ Car & 57443.1342 & SCI & 10 & 30 & 1.315 & 1.09\\
& $\eta$ Car & 57443.1427 & SCI & 10 & 30 & 1.295 & 0.85\\
& HD\,89\,682 & 57443.1646 & CAL & 10 & 30 & 1.175 & 0.87\\
& HD\,89\,682 & 57443.1780 & CAL & 10 & 30 & 1.165 & 0.99\\
& HD\,89\,682 & 57443.1890 & CAL & 10 & 30 & 1.161 & 0.95\\
& $\eta$ Car & 57443.2026 & SCI & 10 & 30 & 1.224 & 0.78\\
& $\eta$ Car & 57443.2142 & SCI & 10 & 30 & 1.222 & 1.05\\
& HD\,89\,682 & 57443.2315 & CAL & 10 & 30 & 1.331 & 1.03\\
& $\eta$ Car & 57443.2689 & SCI & 10 & 30 & 1.263 & 0.77\\
& $\eta$ Car & 57443.2855 & SCI & 10 & 30 & 1.293 & 0.98\\
& HD\,89\,682 & 57443.3078 & CAL & 10 & 30 & 1.350 & 1.16 \\
& HD\,89\,682 & 57443.3692 & CAL & 10 & 30 & 1.706 & - \\
& $\eta$ Car & 57443.3878 & SCI & 10 & 30 & 1.746  & -\\
& $\eta$ Car & 57443.3962 & SCI & 10 & 30 & 1.814 & -\\
& HD\,89\,682 & 57443.4103 & CAL & 10 & 30 & 2.170  & -\\
\hline
27-02-2016 & HD\,89\,682 & 57446.2508 & CAL & 10 & 30 & 1.367  & 1.38 \\
& HD\,89\,682 & 57446.2576 & CAL & 10 & 30 & 1.379  & 1.03 \\
& $\eta$ Car & 57446.2740 & SCI & 10 & 30 & 1.286  & 1.01\\
& $\eta$ Car & 57446.2822 & SCI & 10 & 30 & 1.303 & 0.90\\
& $\eta$ Car & 57446.2911 & SCI & 10 & 30 & 1.325  & 1.06\\
& $\eta$ Car & 57446.2995 & SCI & 10 & 30 & 1.348 & 0.94\\

\hline
\end{tabular}
\end{table*}

\begin{table*}
\caption{$\eta$ Car's 2017 GRAVITY observations}
\label{tab:2017_obs}
\centering
\begin{tabular}{c l l l c c c c} 
\hline\hline
Date & Source & MJD & Type & DIT & NDIT & Airmass & Seeing \\
\hline
30-05-2017 & HD\,89\,682 & 57903.9623 & CAL & 10 & 30 & 1.170 & 0.77 \\
& HD\,89\,682 & 57903.9704 & CAL & 10 & 30 & 1.177 & 1.08 \\
& $\eta$ Car & 57903.9823 & SCI & 10 & 30 & 1.234 & 0.85 \\
& $\eta$ Car & 57903.9905 & SCI & 10 & 30 & 1.241  & 0.80 \\
& $\eta$ Car & 57903.9945 & SCI & 10 & 30 & 1.246 & 0.79 \\
\hline
01-06-2017 & $\eta$ Car & 57906.0526 & SCI & 10 & 30 & 1.385 & 0.55 \\
& $\eta$ Car & 57906.0607 & SCI & 10 & 30 & 1.413  & 0.56 \\
& $\eta$ Car & 57906.0647 & SCI & 10 & 30 & 1.429  & 0.46 \\
& $\eta$ Car & 57906.0728 & SCI & 10 & 30 & 1.462 & 0.61 \\
& $\eta$ Car & 57906.0772 & SCI & 10 & 30 & 1.482 & 0.55  \\
& $\eta$ Car & 57906.0854 & SCI & 10 & 30 & 1.521  & 0.68 \\
& $\eta$ Car & 57906.0894 & SCI & 10 & 30 & 1.542 & 0.58 \\
& HD\,89\,682 & 57906.1007 & CAL & 10 & 30 & 1.693 & 0.56 \\

\hline
\end{tabular}
\end{table*}

\section{Posterior distributions of the best-fit geometrical
  model \label{sec:post_dist}}
\begin{figure*}[hp]
\centering
\begin{minipage}[]{0.45\textwidth}
\includegraphics[width=8.5 cm]{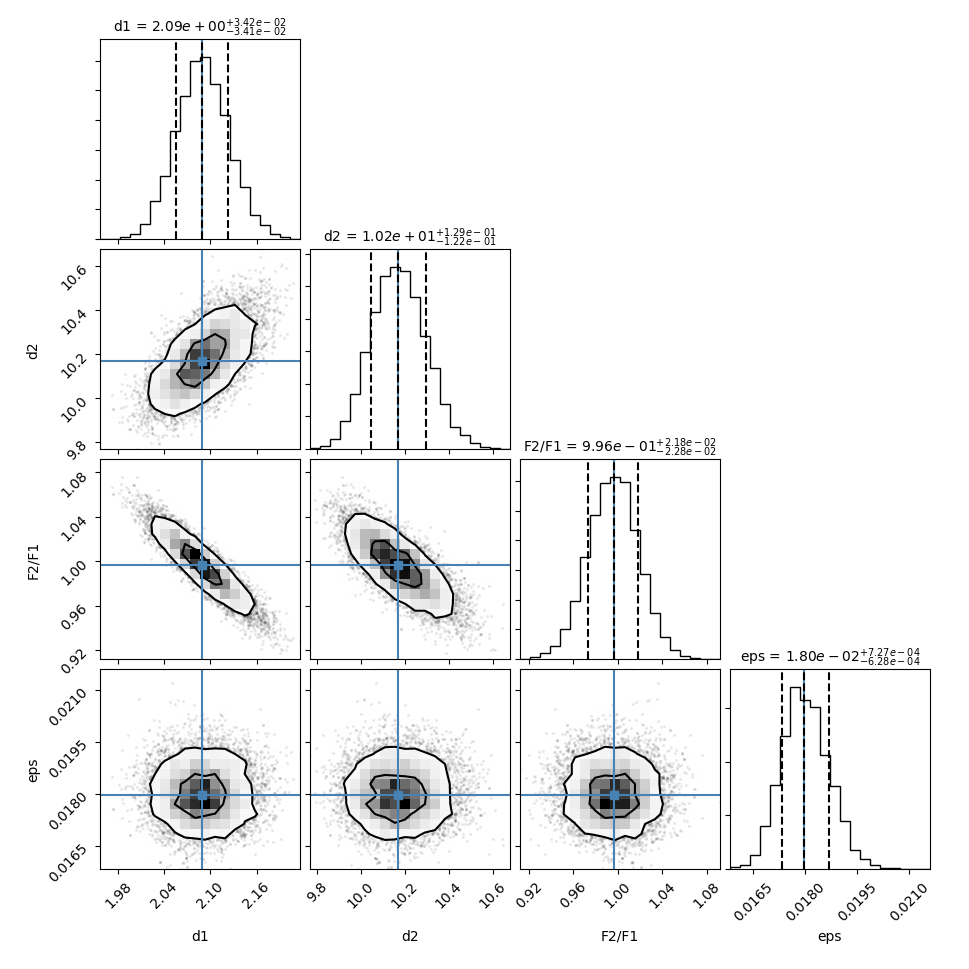}
\end{minipage}
\begin{minipage}[]{0.45\textwidth}
\includegraphics[width=8.5 cm]{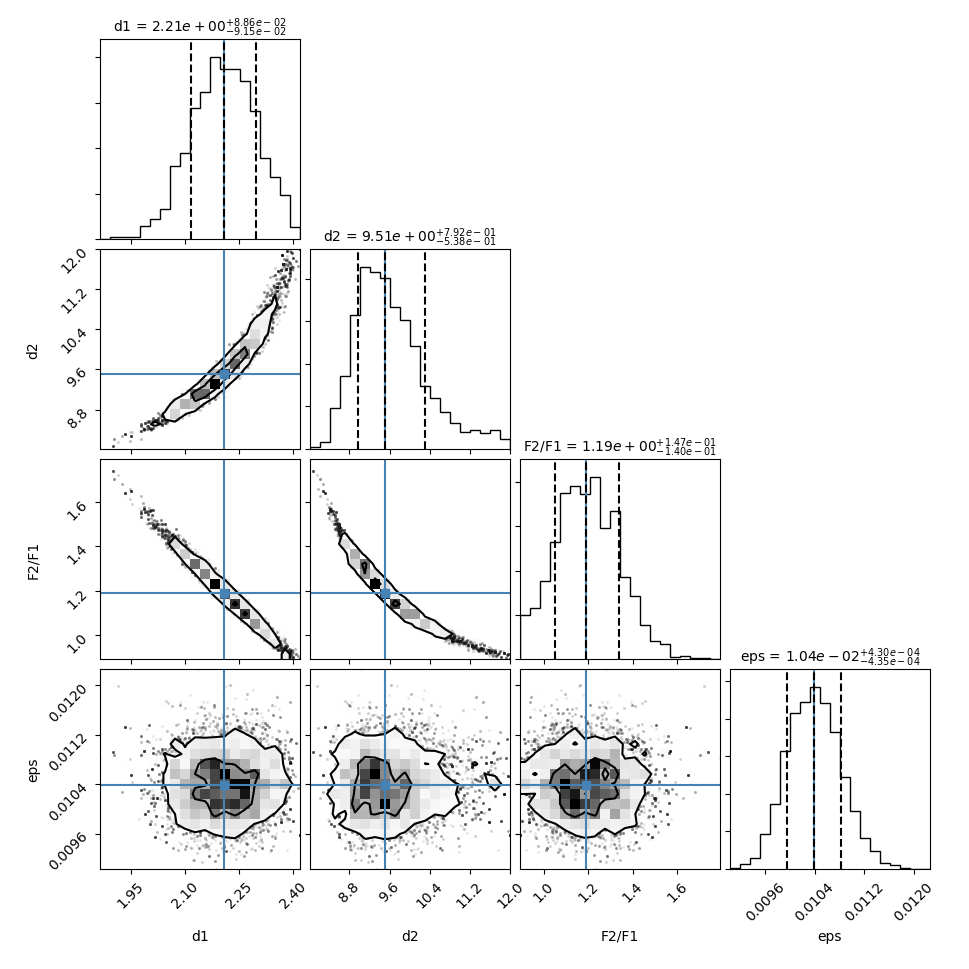}
\end{minipage}

\caption{Posterior distributions of the best-fit parameters of the
  geometrical model presented in Sec.\,\ref{sec:geom_model}. The
  \textit{left} panel shows the posterior distributions for the model
  applied to the V$^2$ data at PA$_{\bot}$ = 40$^{\circ} \pm$
20$^{\circ}$, while the \textit{right} panel displays the posterior
distributions for the model at PA$_{\parallel}$ =  130$^{\circ} \pm$
20$^{\circ}$. The 2D
  distributions show 1 and 2 standard deviations encircled with a
  black contour. The mean of each distribution is
  displayed with a blue square. The 1D histograms show the expected
  value (mean) and $\pm$1$\sigma$ with vertical dashed
lines, together with their corresponding values at the top.}
\label{fig:post_dist} 
\end{figure*}

\section{Dirty beams of the GRAVITY imaging
  epochs \label{sec:dirty_beams}}

\begin{figure*}[htp]
\centering
\includegraphics[width=\textwidth]{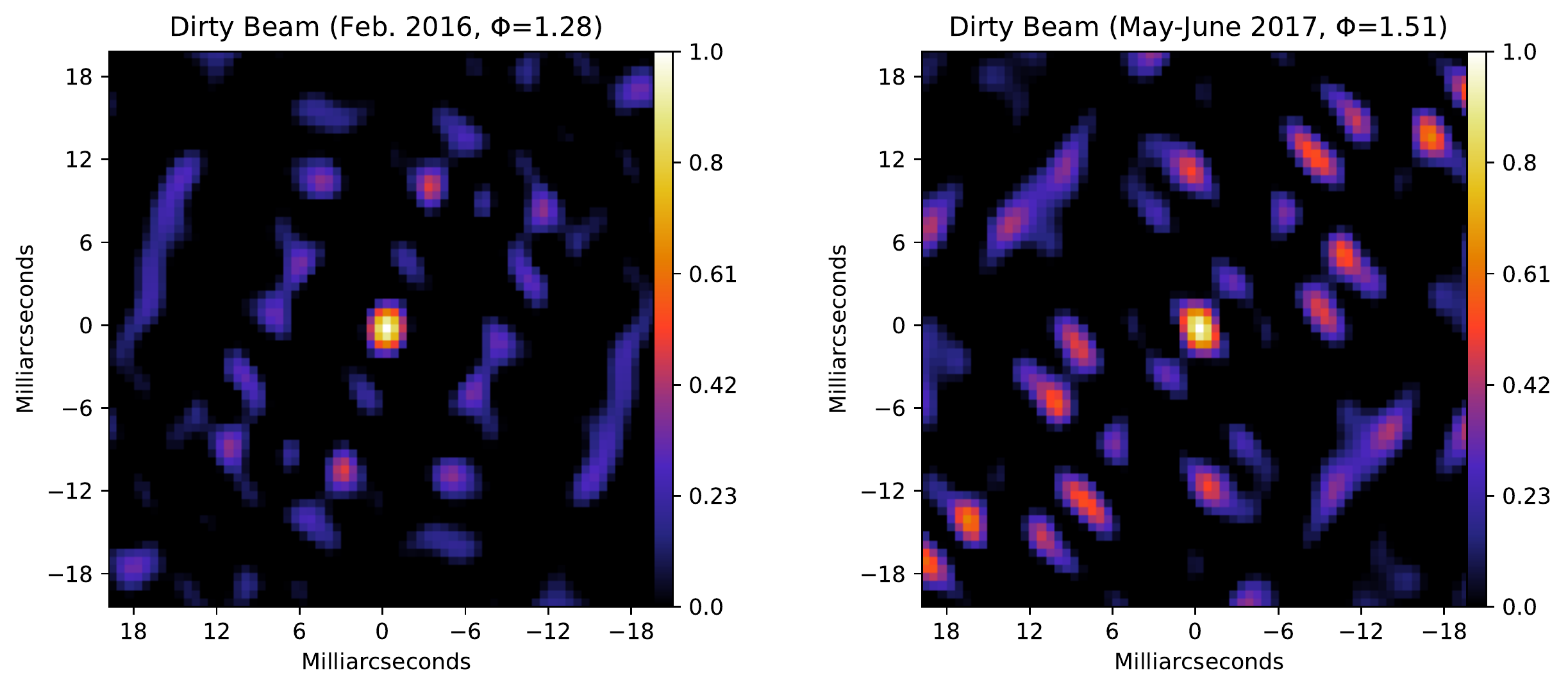}
\caption{Dirty beams of the 2016 (\textit{left}) and 2017 (\textit{right})
GRAVITY data. Notice the strong secondary lobes caused by the more
sparse $u-v$ coverage of the second epoch. }
\label{fig:dirty_beam} 
\end{figure*}

\section{Reconstructed Br$\gamma$ and He {\sc i} images from the 2017
  GRAVITY data \label{sec:2017_ims}}

\begin{figure*}[htp]
\centering
\includegraphics[width=19 cm]{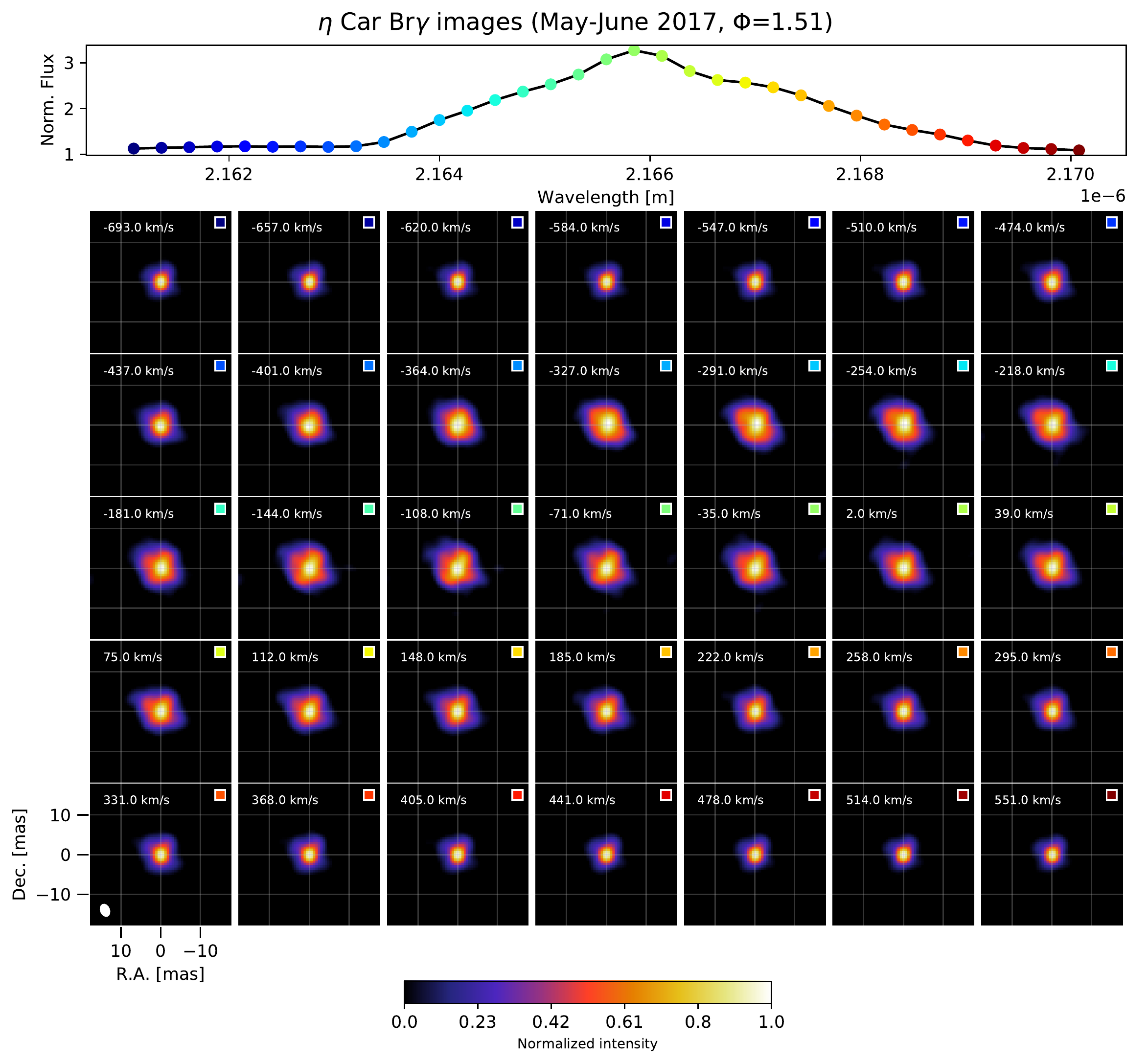}
\caption{Br$\gamma$ interferometric aperture synthesis images from the May-June 2017 data. The
  Doppler velocity of each frame is labeled in the images. For all the panels, east is to the left and north to the
  top and the displayed FOV corresponds to 36$\times$36
  mas. The small white ellipse shown in the lowermost-left panel corresponds
  to the synthesized beam (the detailed PSF is shown in Fig.\,\ref{fig:BrG_p1_june2017}). Above all the images, the GRAVITY
  spectrum is shown and the different positions where the images are
  reconstructed across the line are labeled with a colored square,
  which is also plotted in the images for an easy identification. Due to the sparseness of the $u-v$
  coverage, the quality of these reconstructed images is limited,
  creating a clumpy fine structure and a cross-like shape superimposed
on the source's brightness distribution. Therefore, they cannot be
properly compared with the GRAVITY 2016 data presented in this work. }
\label{fig:BrG_p1_june2017} 
\end{figure*}

\begin{figure*}[htp]
\centering
\includegraphics[width=19 cm]{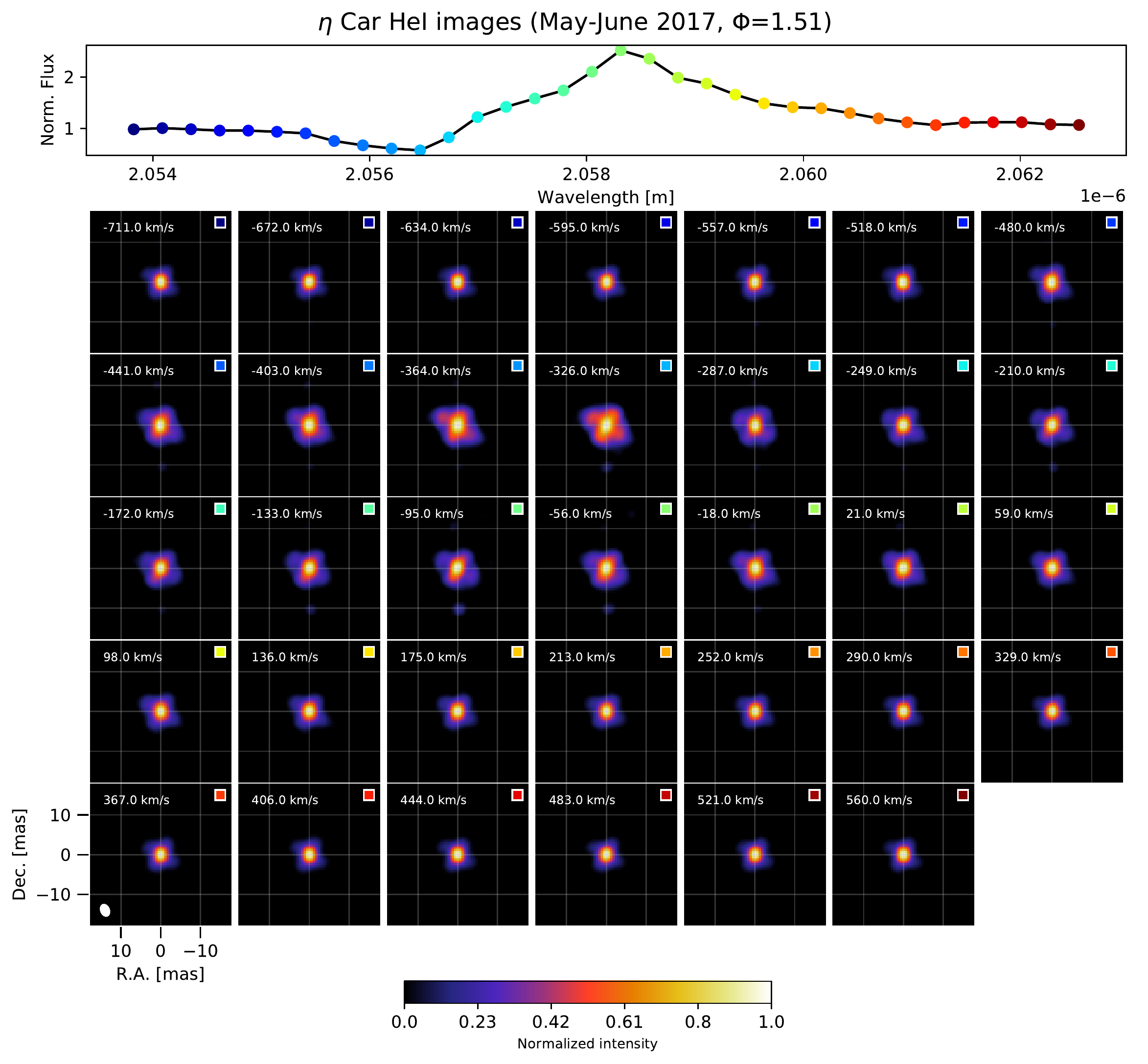}
\caption{He {\sc i} interferometric aperture synthesis images from the May-June 2017
  data. The maps are as described in Figure \ref{fig:BrG_p1_june2017}.}
\label{fig:HeI_p1_june2017} 
\end{figure*}

\section{Best-fit of the reconstructed images to the V$^2$ and
  closure phases \label{sec:fit_v2_cp}}

\begin{figure*}[htp]
\centering
\includegraphics[width=24 cm, angle=90]{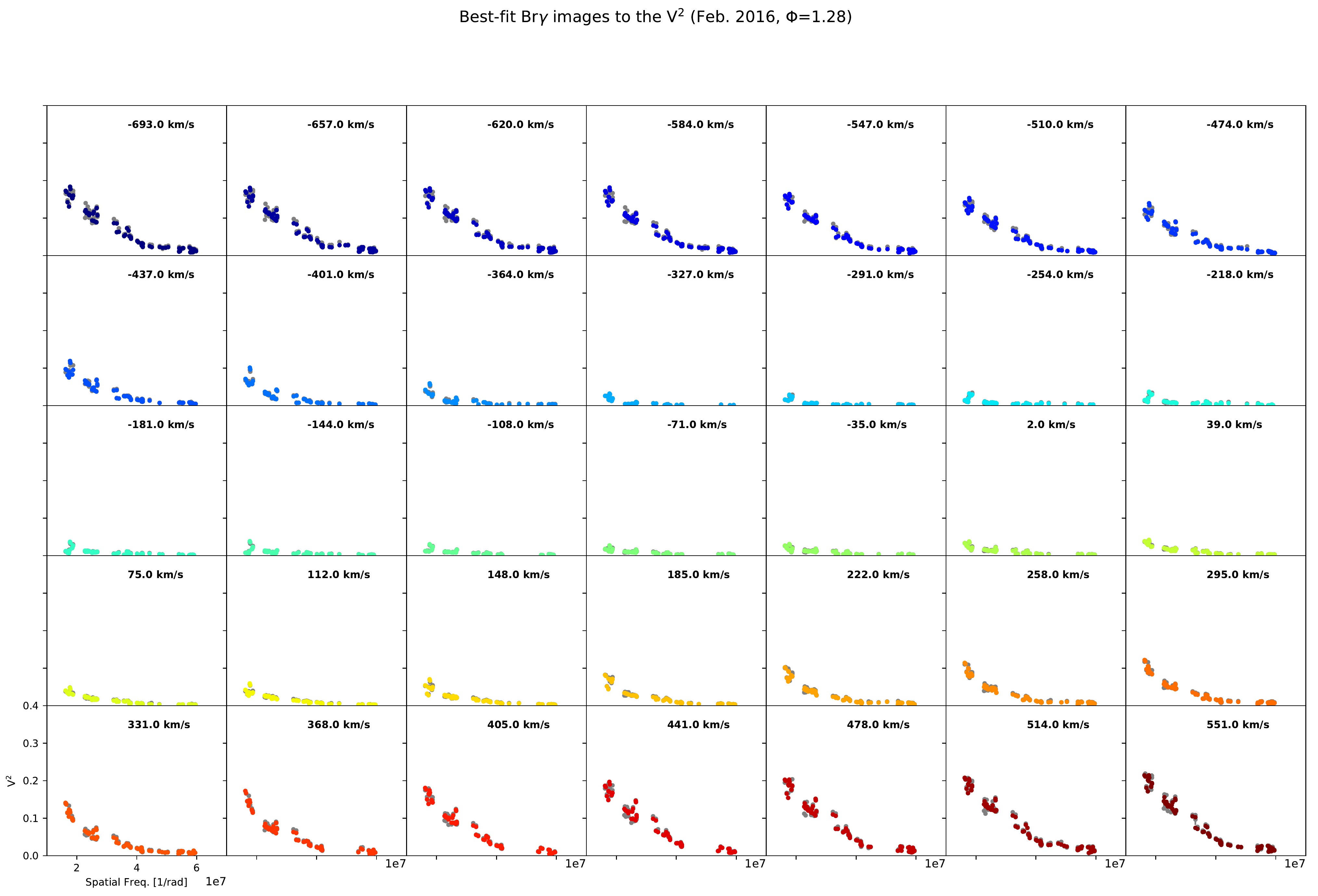}
\caption{Fit to the observed $V^2$ from the interferometric aperture synthesis images across the Br$\gamma$ line
  (2016). The colored dots correspond to the synthetic $V^2$ extracted
  from the
  unconvolved reconstructed images, while the observational data are represented
  with gray dots with 1$\sigma$ error-bars. The
  Doppler velocity is labeled in each panel. All the panels share the
  same horizontal and vertical scales. }
\label{fig:BrG_p1_v2_feb2016} 
\end{figure*}

\begin{figure*}[htp]
\centering
\includegraphics[width=24 cm, angle=90]{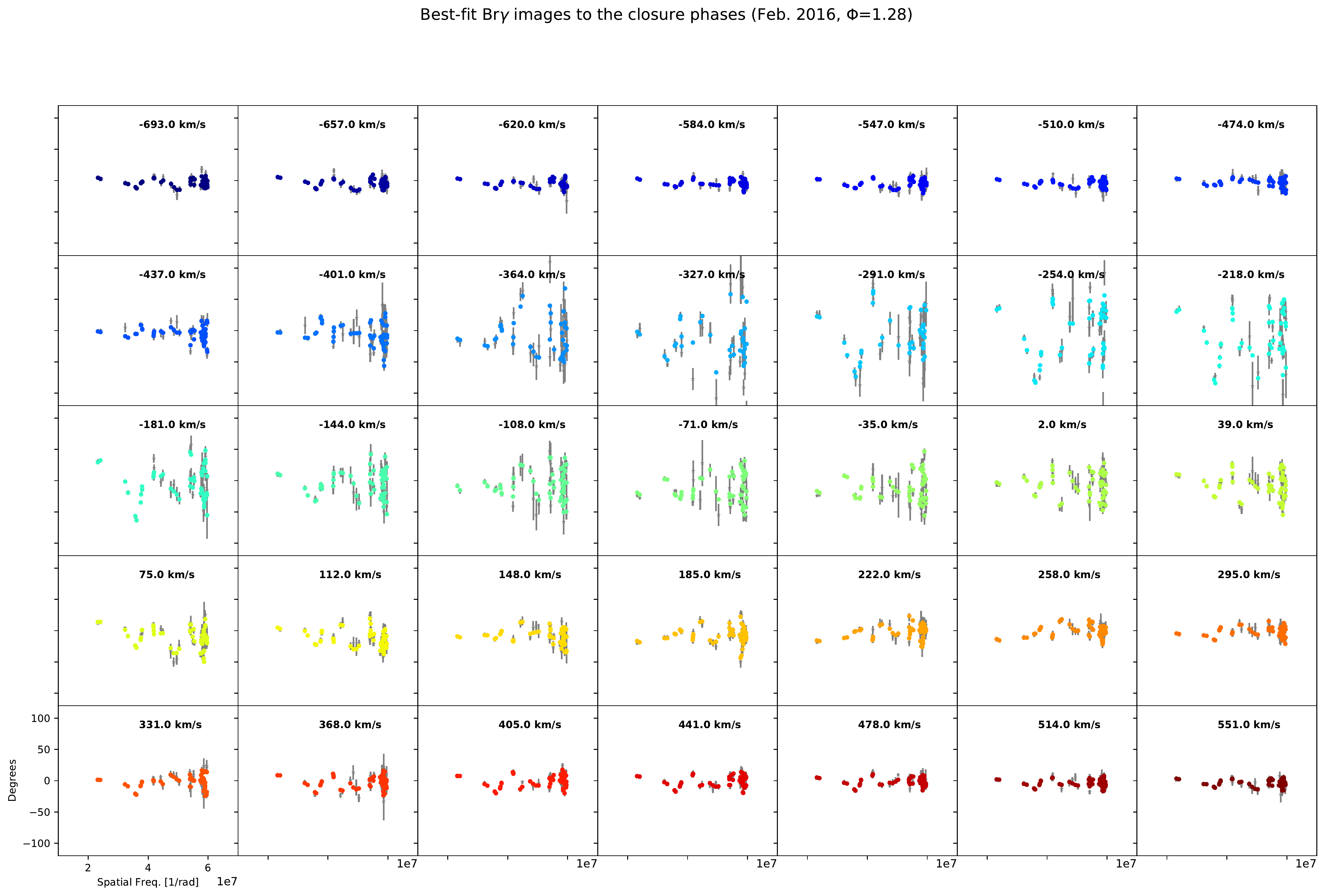}
\caption{Fit to the observed closure phases from the interferometric aperture synthesis images across the Br$\gamma$ line
  (2016). The panels are as described in Figure \ref{fig:BrG_p1_v2_feb2016}. }
\label{fig:BrG_p1_cp_feb2016} 
\end{figure*}

\begin{figure*}[htp]
\centering
\includegraphics[width=24 cm, angle=90]{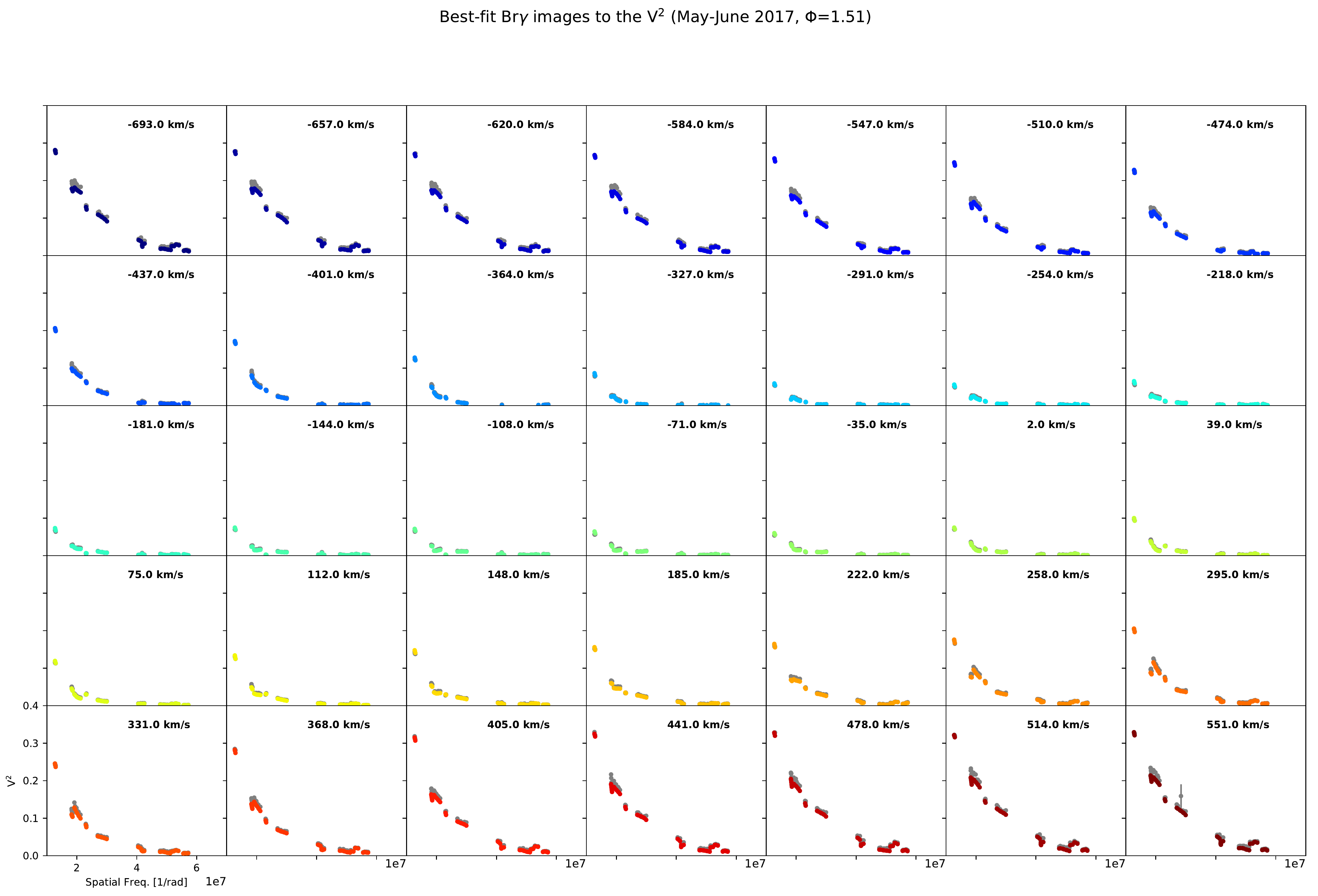}
\caption{Fit to the observed $V^2$ from the interferometric aperture synthesis images across the Br$\gamma$ line
  (2017). The panels are as described in Figure \ref{fig:BrG_p1_v2_feb2016}. }
\label{fig:BrG_p1_v2_june2017} 
\end{figure*}

\begin{figure*}[htp]
\centering
\includegraphics[width=24 cm, angle=90]{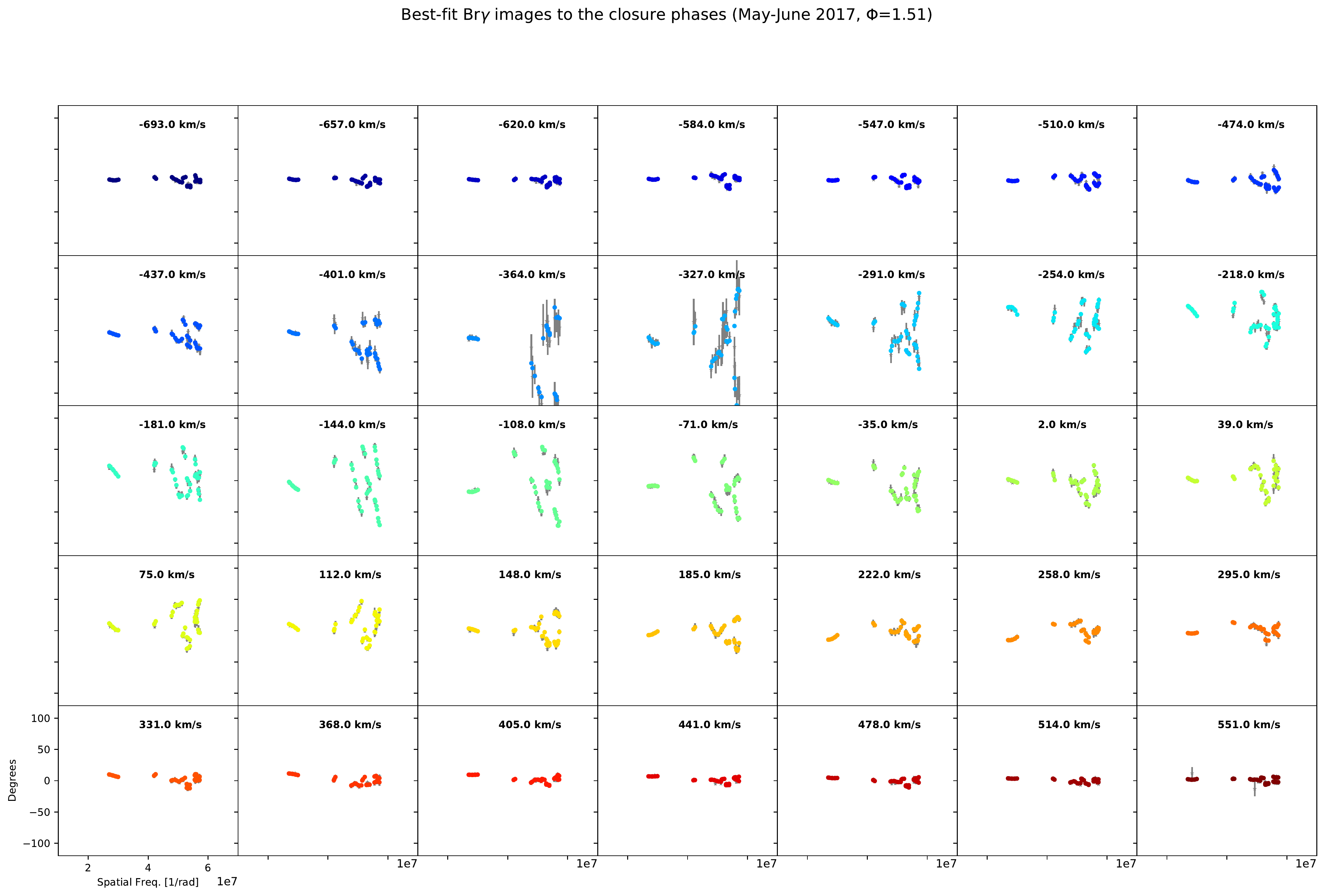}
\caption{Fit to the observed closure phases from the interferometric aperture synthesis images across the Br$\gamma$ line
  (2017). The panels are as described in Figure \ref{fig:BrG_p1_v2_feb2016}. }
\label{fig:BrG_p1_cp_june2017} 
\end{figure*}

\begin{figure*}[htp]
\centering
\includegraphics[width=24 cm, angle=90]{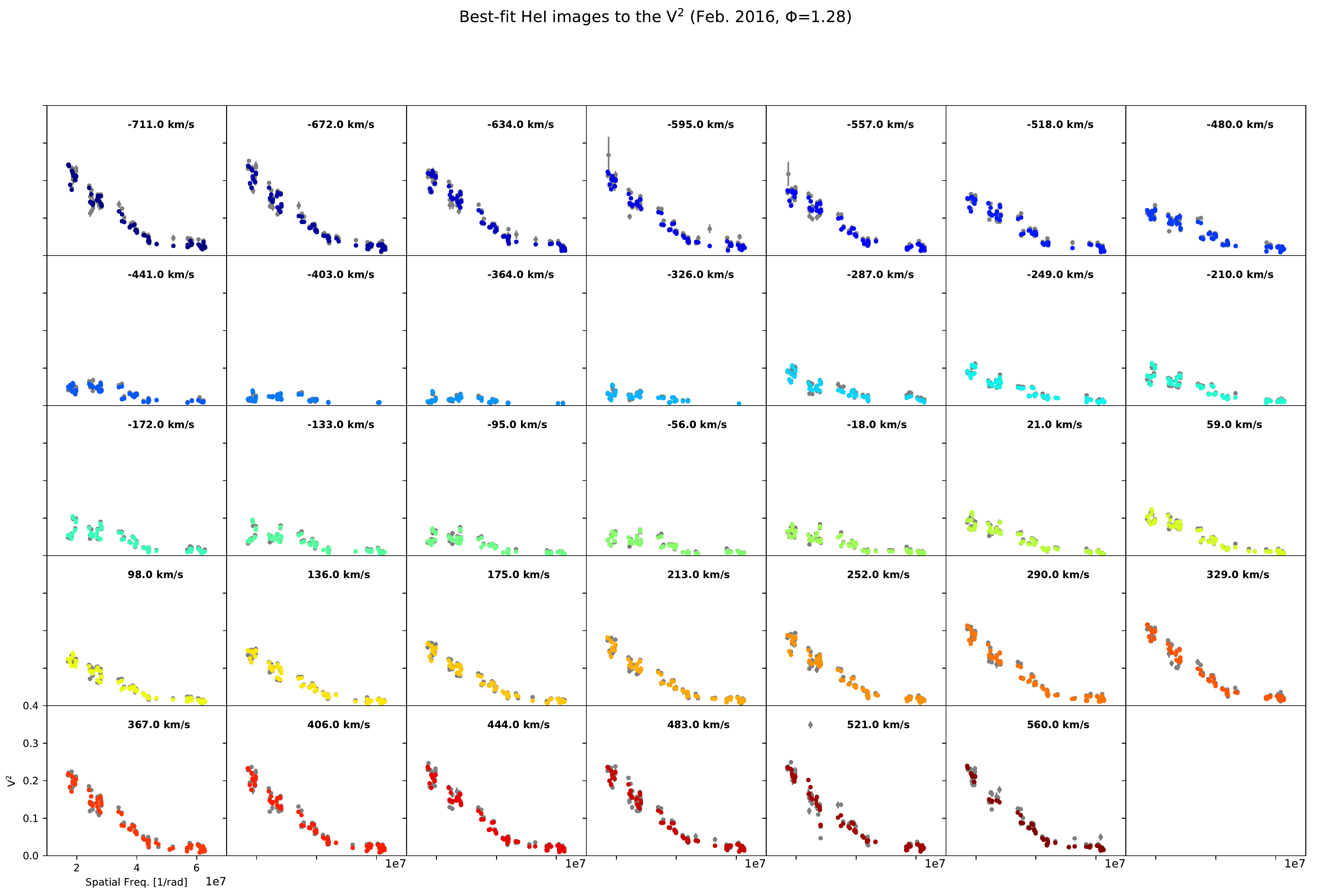}
\caption{Fit to the observed $V^2$ from the interferometric aperture
  synthesis images across the He {\sc i} line
  (2016). The panels are as described in Figure \ref{fig:BrG_p1_v2_feb2016}. }
\label{fig:HeI_p1_v2_feb2016} 
\end{figure*}

\begin{figure*}[htp]
\centering
\includegraphics[width=24 cm, angle=90]{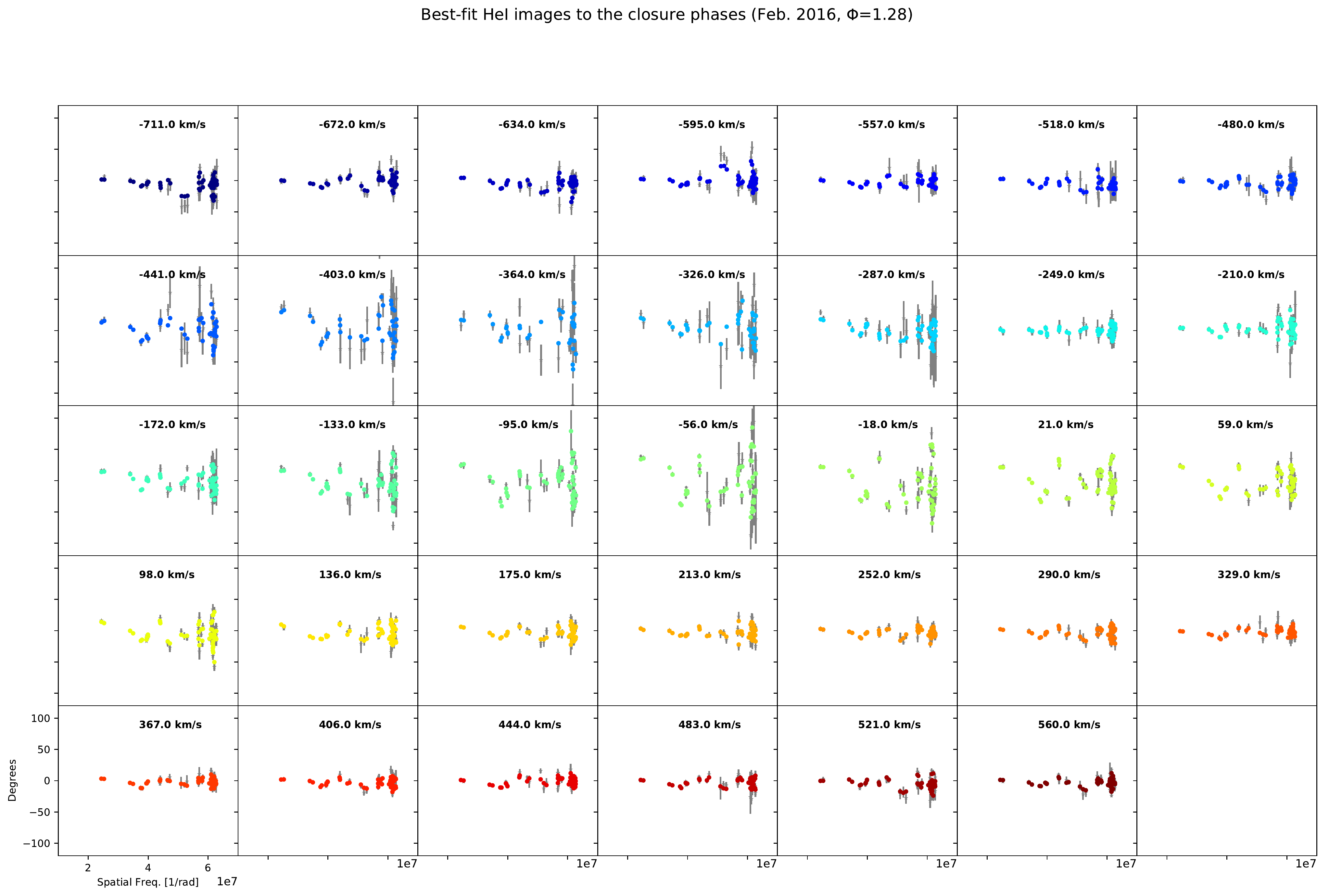}
\caption{Fit to the observed closure phases from the interferometric
  aperture synthesis images across the He {\sc i} line
  (2016). The panels are as described in Figure \ref{fig:BrG_p1_v2_feb2016}. }
\label{fig:HeI_p1_cp_feb2016} 
\end{figure*}

\begin{figure*}[htp]
\centering
\includegraphics[width=24 cm, angle=90]{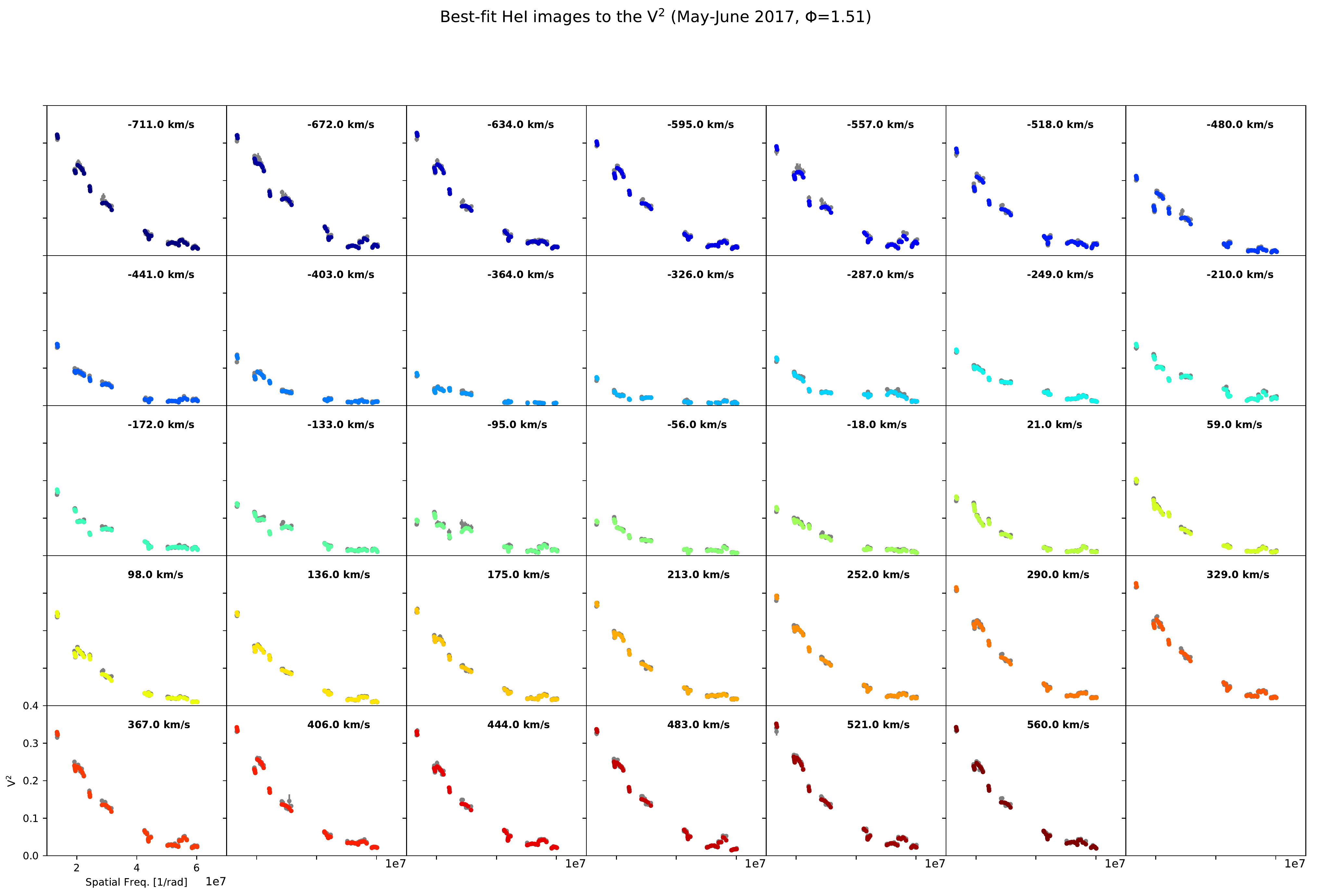}
\caption{Fit to the observed $V^2$ from the interferometric aperture
  synthesis images across the He {\sc i} line
  (2017). The panels are as described in Figure \ref{fig:BrG_p1_v2_feb2016}. }
\label{fig:HeI_p1_v2_june2017} 
\end{figure*}

\begin{figure*}[htp]
\centering
\includegraphics[width=24 cm, angle=90]{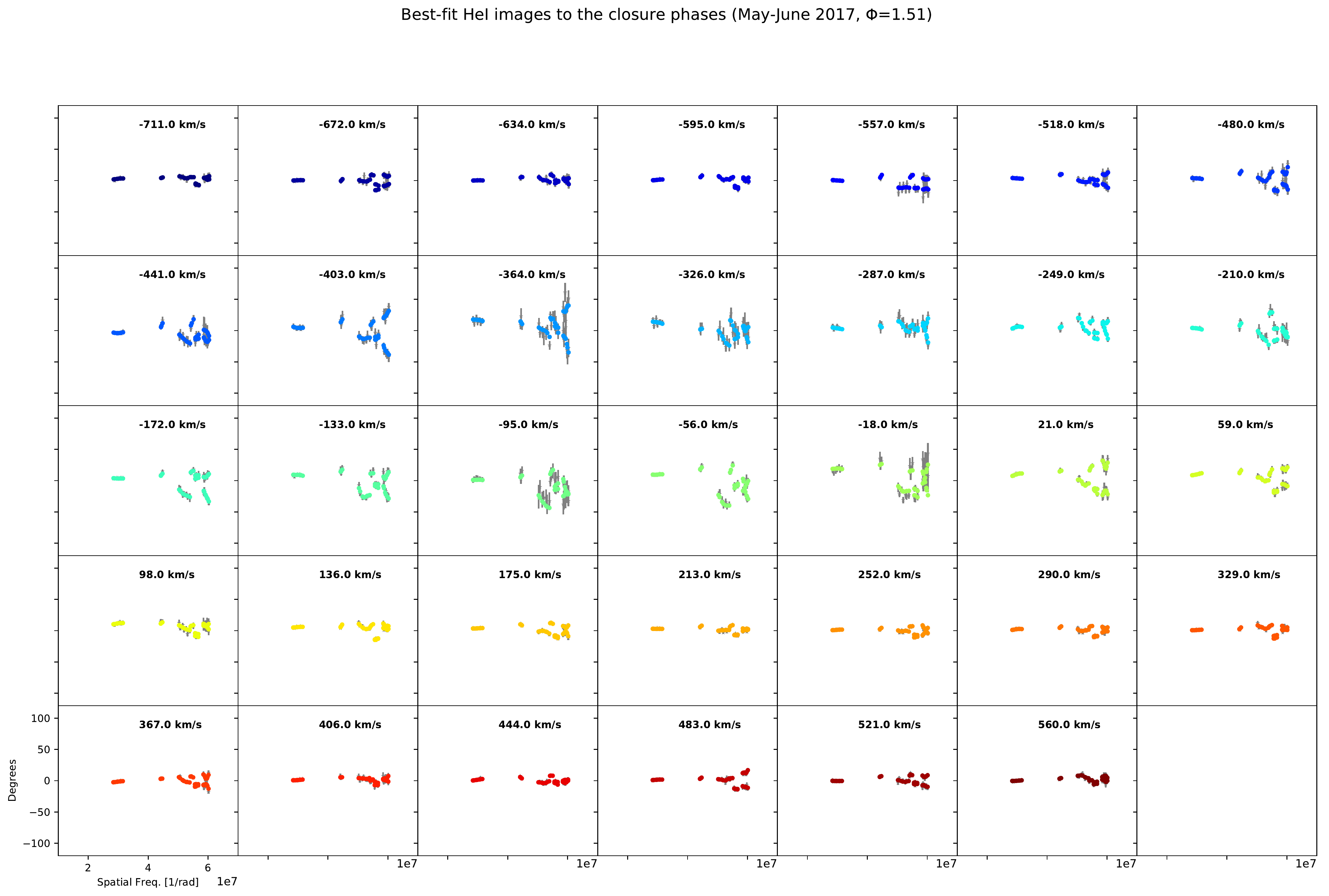}
\caption{Fit to the observed closure phases from the interferometric
  aperture synthesis images across the He {\sc i} line
  (2017). The panels are as described in Figure \ref{fig:BrG_p1_v2_feb2016}. }
\label{fig:HeI_p1_cp_june2017} 
\end{figure*}

\section{\texttt{CMFGEN} models of the FEROS $\eta$ Car spectrum}

\begin{figure*}[htp]
\centering
\includegraphics[width=17 cm]{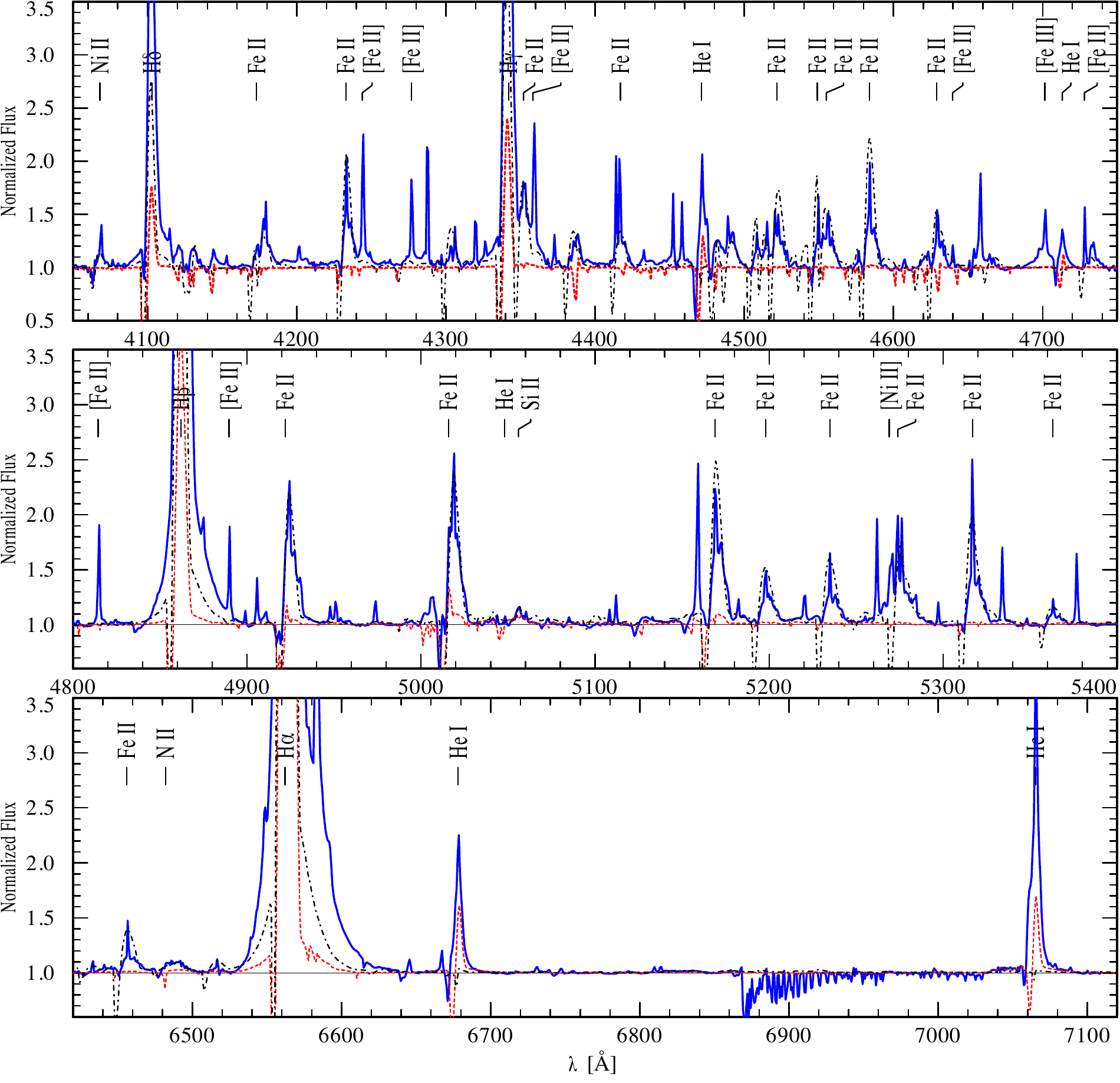}
\caption{ 2016 $\eta$ Car's FEROS spectrum (blue-solid line). The image displays our best-fit
  \texttt{CMFGEN} model (red-dashed line) and the model described in
  \citet{Groh_2012a} (black-dashed line). The Balmer lines of the
  FEROS spectrum are too strong. They include large contributions of
  the circumstellar emission as a result of the bad seeing of
  $\sim$2'' at the time of the observation. The black synthetic
  spectrum matches the Fe {\sc ii} and Si {\sc ii} lines. The
  absorption component of the P-Cygni profiles are filled in and the
  He {\sc i} emission lines are not reproduced by the model, which has
  been already observed and discussed in detail by
  \citet{Hillier_2001}. The hotter red model can hardly fit the
  spectral lines observed at FEROS' wavelengths.
 }
\label{fig:FEROS_spectrum} 
\end{figure*}

\end{appendix}


\end{document}